\documentclass[format=acmsmall, review=false, screen=true]{acmart}
\usepackage{booktabs} % For formal tables

\usepackage[ruled]{algorithm2e} % For algorithms

\SetAlFnt{\small}
\SetAlCapFnt{\small}
\SetAlCapNameFnt{\small}
\SetAlCapHSkip{0pt}
\IncMargin{-\parindent}

\usepackage{xspace}
\usepackage{multibib}
\newcites{sel}{List of Core Papers}

\usepackage{dblfnote}
\DFNalwaysdouble % for this example

\usepackage{color}
\usepackage{xcolor}
\usepackage[normalem]{ulem} % for \sout
\newenvironment{versiontwo}{}{}
\newcommand{\vtwo}[1]{#1}
\newcommand{\vthree}[1]{#1}

\usepackage{ifthen}
\usepackage{amssymb}
\usepackage{adjustbox}   
\newboolean{showcomments}
\setboolean{showcomments}{true} % toggle to show or hide comments
\ifthenelse{\boolean{showcomments}}
  {\newcommand{\nb}[2]{
    \fcolorbox{gray}{yellow}{\bfseries\sffamily\scriptsize#1}
    {\sf\small$\blacktriangleright$\textit{#2}$\blacktriangleleft$}
   }
   
  }
  {\newcommand{\nb}[2]{}
   
  }

\let\oldparagraph\paragraph
\renewcommand{\paragraph}[1]{\textbf{\oldparagraph{#1}}}

\newcommand\devops{DevOps\xspace}
\newcommand\numberofselectedpapers{167\xspace}
\newcommand\numberofmainselectedpapers{50\xspace}
\newcommand\pageofmainselecetdstudies{31\xspace}
\newcommand{\myparagraph}[1]{\vspace{.5em}\noindent\textit{\textbf{#1:}}\hspace{.3em}}
\newcommand{\swurl}[1]{\footnote{\url{#1}}}

\acmJournal{CSUR}
\acmYear{2019}
 \copyrightyear{2019}
\setcopyright{othergov}
\acmYear{2019} \acmVolume{52} \acmNumber{6} \acmArticle{127} \acmMonth{11} %\acmPrice{15.00}
\acmDOI{10.1145/3359981}

\acmDOI{10.1145/3359981}

\received{January 2019}
\received[revised]{June 2019}
\received[accepted]{August 2019}

\begin{document}
% Title portion. Note the short title for running heads
\title{A Survey of DevOps Concepts and Challenges}

\author{Leonardo Leite}
%\orcid{1234-5678-9012-3456}
\affiliation{%
  \institution{University of S\~ao Paulo}
%  \department{Department of Computer Science}
%  \streetaddress{Rua do Matao 1010}
  \city{S\~ao Paulo}
%  \state{SP}
%  \postcode{05508-090}
  \country{Brazil}}
\email{leofl@ime.usp.br}

\author{Carla Rocha}
%\orcid{1234-5678-9012-3456}
\affiliation{%
  \institution{University of Bras\'ilia}
%  \department{Department of Computer Science}
%  \streetaddress{Rua do Matao 1010}
  \city{Brasilia}
%  \state{SP}
%  \postcode{05508-090}
  \country{Brazil}}
\email{caguiar@unb.br}

\author{Fabio Kon}
%\orcid{1234-5678-9012-3456}
\affiliation{%
  \institution{University of S\~ao Paulo}
%  \department{Department of Computer Science}
%  \streetaddress{Rua do Matao 1010}
  \city{S\~ao Paulo}
%  \state{SP}
%  \postcode{05508-090}
  \country{Brazil}}
\email{kon@ime.usp.br}

\author{Dejan Milojicic}
%\orcid{1234-5678-9012-3456}
\affiliation{%
  \institution{Hewlett Packard Labs, Palo Alto}
%  \department{Department of Computer Science}
%  \streetaddress{Rua do Matao 1010}
  \city{Palo Alto}
%  \state{SP}
%  \postcode{05508-090}
  \country{USA}}
\email{dejan.milojicic@hpe.com}

\author{Paulo Meirelles}
%\orcid{1234-5678-9012-3456}
\affiliation{%
  \institution{Federal University of S\~ao Paulo}
%  \department{Department of Computer Science}
%  \streetaddress{Rua do Matao 1010}
  \city{S\~ao Paulo}
%  \state{SP}
%  \postcode{05508-090}
  \country{Brazil}}
\email{paulo.meirelles@unifesp.br}

\begin{abstract}
\devops is a collaborative and multidisciplinary organizational effort to automate continuous delivery of new software updates while guaranteeing their correctness and reliability.
The present survey investigates and discusses DevOps challenges from the perspective of engineers, managers, and researchers.  We review the literature and develop a DevOps conceptual map, correlating the DevOps automation tools with these concepts. We then discuss their practical implications for engineers, managers, and researchers. Finally, we critically explore some of the most relevant DevOps challenges reported by the literature. 
\end{abstract}

%
% The code below should be generated by the tool at
% http://dl.acm.org/ccs.cfm
% Please copy and paste the code instead of the example below.
%
% Tree https://dl.acm.org/ccs/ccs.cfm
% Instructions: https://www.acm.org/binaries/content/assets/publications/article-templates/ccs-howto-v6-12jan2015.docx
% Relevance scores: 500 - high; 300 - medium; 100 - low
% Use any first- and second-level nodes of the classification scheme which are relevant to your paper, then look at the lower levels under them. 
% Identify the lowest branches of the tree that seem to apply to your particular paper. Use the lowest node in the tree available.
% It is not uncommon to include 6 or more terms for an article.


%
% End generated code
%


\maketitle

% The default list of authors is too long for headers.
\renewcommand{\shortauthors}{L. Leite et al.}

% English help: https://corpus.byu.edu/coca/

%\info{Paper length: should not exceed 35 pages}
\section{Introduction}
\label{sec:intro} 

Even though the DevOps movement has been discussed for nearly a decade, it lacks a widely accepted definition. By consolidating the most cited definitions of DevOps, we crafted our definition, similar to the one proposed by Dyck \emph{et al.}~\citesel{dyck2015release}, which we adopt throughout this paper:

\begin{quotation}
\emph{DevOps is a collaborative and multidisciplinary effort within an organization 
to automate continuous delivery of new software versions, 
while guaranteeing their correctness and reliability.}
\end{quotation}

Desiring to improve their delivery process~\cite{puppet2014devops}, enterprises are widely adopting \devops~\citesel{chen2015huge,gray2006conversation,feitelson2013facebook}. \vtwo{Although in discordance with most of the academic definitions, the software industry also uses the word ``DevOps'' to describe a well-paid job title~\citesel{hussain2017zealand}\cite{stackoverflow2017earn}.} Becoming a \devops engineer is an attractive opportunity for software professionals. \devops is also an important phenomenon studied by software engineering researchers and already a mandatory topic in software engineering courses. 

\devops and its \textbf{challenges} can be discussed from three perspectives: engineers, managers, and researchers. \textbf{Engineers} benefit from \textbf{(1)} qualifying themselves for a \devops position, a technically hard task guided by over 200 papers, 230 books, and 100 tools~\cite{xebia2016table}. Engineers also need to \textbf{(2)} learn how to re-architect their systems to embrace continuous delivery. \textbf{Managers} want to know \textbf{(3)} how to introduce \devops into an organization and \textbf{(4)} how to assess the quality of already-adopted \devops practices.  Managers and engineers, also referred to as \emph{practitioners}, share the necessity of choosing the best automation toolset. Finally, academic \textbf{researchers}  \textbf{(5)} conduct studies to determine the state of practice in \devops, thereby contributing to discussions among engineers and managers, and \textbf{(6)} educate a new generation of software engineers on \devops principles and practices.

\begin{versiontwo}
In this context, our \emph{research problem} is devising a conceptual framework to guide engineers, managers, and academics in the exploration of DevOps tools, implications, and challenges. 
There are several studies and a few surveys tackling \devops challenges. However, with few exceptions~\citesel{neely2013easy}, they focus on a single perspective to address a given problem. 
In this context, this survey contributes to the field by investigating and discussing \devops concepts and challenges from multiple perspectives: engineers, managers, and researchers. Our review also explores a much broader range of sources and is more up-to-date than previous studies. We exploit technical implication and complexity of adopting \devops such as automation, tightly versus loosely coupled architectures, containerization versus configuration management to deployment automation, and toolset management. Previous surveys did not cover these practical aspects and implications of \devops. 
\end{versiontwo}

We first survey and analyze an academic bibliography. We explain the selection and analysis procedures of these works in Section~\ref{sec:methodology}. The selected studies are categorized and described in Section~\ref{sec:sources}. Then, we complement our discussions with non-academic sources, such as books and posts from practitioners' blogs; we describe these other sources in Section~\ref{sec:sources}. Also in this section, we compare our work to other reviews on \devops.

\vtwo{Based on the reviewed literature, Section~\ref{sec:concepts} presents the construction of a conceptual framework~\cite{miles1994qualitative} on DevOps composed of five conceptual maps.} We relate these concepts to the engineer and manager perspectives.  We also look at the practical side of \devops by analyzing the \devops tools under the perspective of the concepts in Section~\ref{sec:tools}. We categorize these tools, correlate them to concepts, and discuss which roles in the organization should use which tools.

The discussion of \devops concepts and tools lays the basis to raise essential implications for engineers, managers, and researchers, which we present in Section~\ref{sec:implications}. Although most of the implications are straightforward, some challenges are not entirely illuminated by the current literature. We summarize and discuss some of the main \devops challenges in Section~\ref{sec:discussion}, debating limitations and even contradictions found in the literature.

We close with the limitations of this survey in Section~\ref{sec:limitations} and with concluding remarks in Section~\ref{sec:conclusions}. In the next section, we briefly introduce \devops history, motivations, and goals.

%%%%%%%%%%%%%%%%%%%%%%%%%%%%%%
\section{DevOps}
\label{sec:devops} 

\devops is an evolution of the agile movement~\citesel{christensen2016teaching}. 
Agile Software Development advocates small release iterations with customer reviews. It assumes the team can release software frequently in some production-like environment. However, pioneer Agile literature puts little emphasis on deployment-specific practices.
No Extreme Programming (XP) practice, for example, is about deployment~\cite{beck2004xp}.
The same sparse attention to deployment is evident in traditional software engineering processes~\cite{kroll2003rup} and books~\cite{pressman2005software,sommerville2011software}. Consequently, the transition to production tends to be a stressful process in organizations, containing manual, error-prone activities, and, even, last-minute corrections~\cite{humble2010continuous}.
\devops proposes a complementary set of agile practices to enable the iterative delivery of software in short cycles effectively.

From an organizational perspective, the DevOps movement promotes closer collaboration between developers and operators. The existence of distinct silos for operations and development is still prevalent: operations staff are responsible for managing software modifications in production and for service levels~\citesel{woods2016view}; development teams, on the other hand, are accountable for continuously developing new features to meet business requirements. Each one of these departments has its independent processes, tools, and knowledge bases. The interface between them in the pre-DevOps era was usually a ticket system: development teams demanded the deployment of new software versions, and the operations staff manually managed those tickets. 

In such an arrangement, development teams continuously seek to push new versions into production, while operations staff attempt to block these changes to maintain software stability and other non-functional concerns. Theoretically, this structure provides higher stability for software in production. However, in practice, it also results in long delays between code updates and deployment, as well as ineffective problem-solving processes, in which organizational silos blame each other for production problems.

Conflicts between developers and operators, significant deployment times, and the need for frequent and reliable releases led to inefficient execution of agile processes. In this context, developers and operators began collaborating within enterprises to address this gap. This movement was coined ``DevOps'' in 2008~\cite{debois2008devops}.

In the \emph{Continuous Delivery} book~\cite{humble2010continuous}, Humble advocates for an \emph{automated deployment pipeline}, in which any software version committed to the repository must be a production-candidate version. After passing through stages, such as compilation and automated tests, the software is sent to production by the press of a button. This process is called \emph{Continuous Delivery}. A variant is the \emph{continuous deployment}~\cite{humble2010deployment}, which automatically sends to production every version that passes through the pipeline. Many authors closely relate \devops to continuous delivery and deployment~\citesel{shahin2016architecting,wettinger2015environments,yasar2016security,wettinger2015knowledge}. Yasar, for example, calls the deployment pipeline a ``DevOps platform''~\citesel{yasar2016security}.

Besides automating the delivery process, \devops initiatives have also focused on using automated runtime monitoring for improving software runtime properties, such as performance, scalability, availability, and resilience~\citesel{christensen2016teaching,basiri2016chaos,balalaie2016microservices,roche2013quality}. From this perspective, ``Site Reliability Engineering''~\cite{beyer2016site} emerged as a new term related to \devops work at runtime.

\section{Study design}
\label{sec:methodology}

\begin{figure}[ht]
 \centering
  \includegraphics[scale=0.85]{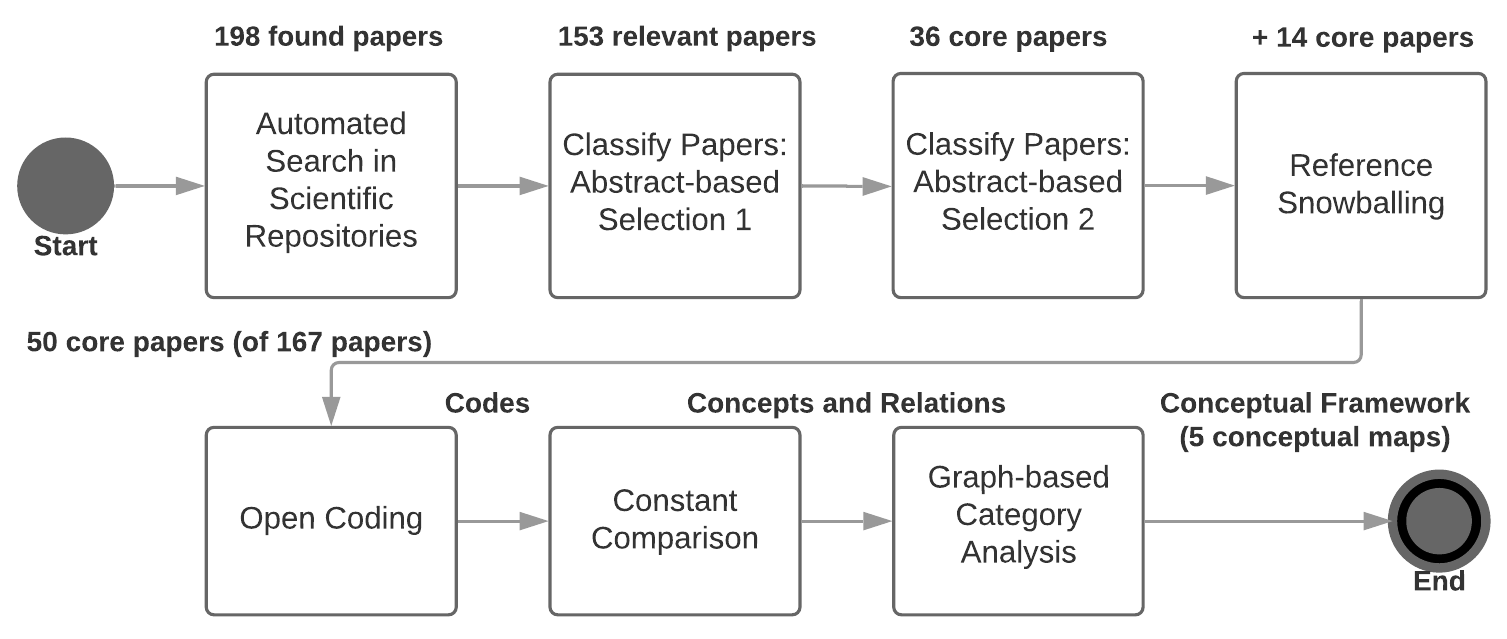}
  \caption{\vthree{Sequence of steps taken in our study}}
  \label{fig:study-design}
\end{figure}

The recent growth of the DevOps community prompted a rapid increase in publications, imposing a barrier to a thorough analysis of every published work on the matter. We adopted a qualitative procedure to limit the number of analyzed studies and the information extracted from them. We established a search and selection protocol, inspired by the procedures of systematic literature reviews~\cite{budgen2006review}, and set analysis procedures based on Grounded Theory strategies~\cite{charmaz2008grounded}. \vtwo{Our primary outcome is the construction of a conceptual framework, composed of a set of conceptual maps that describe the main concepts of \devops and how they relate to each other (Section~\ref{sec:concepts}).} In this section, we detail our search and selection protocol, as well as our procedures to build our conceptual maps. \vthree{This sequence is outlined in Figure~\ref{fig:study-design}}.

\subsection{Search and selection protocol}

We searched for articles in major scientific papers repositories, namely: ACM Digital Library; IEEE Xplore Digital Library; the Spring Link. We also explored Google Scholar, but its outcomes did not provide additional relevant results. The search was conducted according to the following criteria:

\begin{itemize}
\item Title or abstract must contain the word ``\devops.'' 
\item Papers must be published in journals, conferences, congresses, or symposiums (workshop papers were not considered). 
\item Papers must be written in English. 
\item Papers must have at least three pages. 
\item Literature reviews were not included. We present them as related work in Section~\ref{sec:related-surveys}. 
\end{itemize}

\begin{versiontwo}

This search resulted in 198 papers. 
We classified these papers into the following categories: irrelevant, relevant, and core papers.
To perform the classification process, we read the abstract of all 198 papers and, in some cases, read their content too.
Irrelevant papers used the keyword DevOps only for their initial background, not being indeed papers about DevOps. Using this criterion, we discarded  45 irrelevant papers, reducing to 153 the number of \vthree{relevant} papers to be analyzed.  \vthree{Then, we applied a second filter employing another set of criteria to identify papers that would be unlikely to generate any further concepts, handling theoretical saturation \cite{coleman2008}. For example, we identified papers with overlapping topics, that would only reinforce already existing concepts.} We then proceeded with our selection process and, \vthree{from the 153 relevant papers}, we obtained a subset of 36 core papers, which were thoroughly read. The remaining 117 \vthree{ones} were classified \vthree{just} as relevant papers. 

 To reduced the subjectivity of this selection process, at least two authors had to agree on each outcome; the authors produced their classifications independently and when there were different classifications, each paper was individually examined until a consensus was reached. Moreover, even though there were no hard rules for the selection process, we followed some guidelines:

\end{versiontwo}

\begin{itemize}
\item If a paper discusses \devops peculiarities or challenges in some specific context, it is a core paper.
\item If a paper is primarily centered on a single tool and does not discuss \devops itself, it is classified as relevant.
\item If a paper has an extended or more cited version, only the most relevant version is a core paper, whereas the other one is just a relevant paper.
\item If several papers are about the same subject, only some of them are selected as core papers, whereas the others are classified as relevant.
\item If a paper is about applying \devops in a specific context, but does not elaborate beyond the state-of-the-art, it is discarded.
\item If a paper discusses some topic (e.g., big data) and uses \devops only as background, it is discarded.
\item If a paper only presents a simple proposal or opinion, without validation, it is discarded.
\item In exceptional circumstances, studies become core papers without following all the above rules if at least three authors agree.
\end{itemize}

We also applied a snowballing process~\cite{claes2014snowballing} to cover important work not found with the query string ``devops''. We explored historical, seminal, highly-cited, and recent references. From this criterion, we considered 14 additional core papers \vthree{(i.e., equaling 167 relevant papers with 50 core papers)}.

We list all the \numberofmainselectedpapers core papers on Page \pageofmainselecetdstudies at the end of this paper. We identify each paper with a unique reference code, composed of a sequence number and a letter indicating how we found the article: ``A'' for the ACM Digital Library; ``I'' for the IEEE Xplore Digital Library; ``S'' for Springer Link; and ``B'' for the snowballing process. This reference code is later used to display the papers on the conceptual maps.

%Tools like \texttt{Mendeley}, \texttt{Libre Office spreadsheets}, \texttt{SQLite} databases, and some Python scripts, like the ones that generated our bar plots using \texttt{Pandas} and \texttt{Pyplot}.

%TODO: Leo revisar/comparar
To support our searching process and the analysis of the papers, we used some tools such as \texttt{Mendeley}, \texttt{Libre Office spreadsheets}, and \texttt{SQLite} database. Moreover, we developed Python scripts, for example, that generated the bar plots presented in Section~\ref{sec:sources}, using \texttt{Pandas} and \texttt{Pyplot}.

\subsection{Producing the conceptual framework}

\vtwo{According to Miles and Huberman, a conceptual framework explains graphically the main things to be studied in a given field, including elements such as key factors and their relationships~\cite{miles1994qualitative}. Our conceptual framework is composed of conceptual maps, which are are diagrams structured as graphs in which nodes depict concepts and arrows represent relationships among concepts~\cite{hager1997designing}.} Our process of building the conceptual maps was based on Grounded Theory, which is a recognized method for conducting emergent qualitative research~\cite{corbin2014slr} and whose adoption in software engineering has grown considerably in recent years~\cite{stol2016grounded}. According to Charmaz, it reduces preconceived ideas about the research problem and helps researchers remain amenable to alternative interpretations and apprehend the data in different ways~\cite{charmaz2008grounded}. Grounded Theory is usually applied to primary sources, such as interviews and observational field-notes~\cite{sbaraini2011grounded}. In this paper, we adopted Grounded Theory strategies in the literature review to handle the inherent subjectivity of conceptual analysis.

A core strategy of Grounded Theory is \emph{coding}~\cite{charmaz2008grounded}, which breaks down and labels data into smaller components~\cite{sbaraini2011grounded}. In the \emph{initial coding}, also called \emph{open coding}, researchers avoid injecting previous assumptions, biases, and motivations~\cite{stol2016grounded}. Researches then apply \emph{constant comparison} by constantly comparing and correlating codes from different sources to produce \emph{concepts} and group concepts into \emph{categories}~\cite{sbaraini2011grounded, stol2016grounded}. 
In our study, we read the \numberofmainselectedpapers core studies and applied \emph{open coding} on them to produce codes. Then, through \emph{constant comparison}, we analyzed, abstracted, and grouped the codes into concepts and categories, presenting each category in a conceptual map. 

\begin{versiontwo}

The open coding phase was carried as follows: while reading a core paper, the researchers used the Mendeley software to highlight excerpts containing potential DevOps concepts. \vthree{In the following, we provide six examples of codes and their corresponding highlighted text}.

{\small

\myparagraph{DevOps emphasizes collaboration}
``DevOps is a new phenomenon in software engineering, emphasizing collaboration (\dots) that bridge software development and operations activities (...)'' \citesel{lwakatare2016towards}. 

\myparagraph{DevOps advocates collaboration}
``DevOps is a set of practices that advocate the collaboration
between software developers and IT operations (...)'' \citesel{claps2015journey}.

\myparagraph{DevOps emphasizes collaboration and communication}
``DevOps is a culture, movement or practice that emphasizes the collaboration and communication of both software developers and Information Technology (IT) professionals (...)'' \citesel{chung2016ksa}.

\myparagraph{DevOps shares knowledge and tools}
``DevOps was introduced to include cross-functional roles of teams with four perspectives: (...), and 4) sharing of knowledge and tools (...)'' \citesel{chung2016ksa}. 

\myparagraph{DevOps has a culture of collaboration / Collaboration means sharing knowledge and tools}
``Achieving these benefits within enterprises requires the discipline of DevOps: a culture of collaboration between all team members; (...); sharing of knowledge and tools (...)'' \citesel{humble2011devops}.

\myparagraph{DevOps emphasizes working together / working together by sharing tools, processes, and practices}
``That is why the DevOps movement's recent emergence is so heartening. It emphasizes development and operations staff working together as early as possible -- sharing tools, processes, and practices to smooth the path to production (...)'' \citesel{woods2016view}.
}

By \vthree{applying constant comparison in} these codes  coming from different sources, we defined the concept ``Culture of collaboration'' and linked it to the ``DevOps'' concept in the form ``(DevOps) $\rightarrow$ is a $\rightarrow$ (culture of collaboration),'' with concepts represented within parenthesis. From the excerpts we also understood that a ``culture of collaboration'' means sharing ``knowledge, tools, processes, and practices'', so we grouped these last concepts,  from different sources, to state that ``(DevOps) $\rightarrow$ is a $\rightarrow$ (culture of collaboration) $\rightarrow$ based on sharing $\rightarrow$ (knowledge, tools, processes, and practices),'' as we can read from Figure~\ref{fig:concepts-people} in Section \ref{sec:concepts}.

As concepts emerged, we assembled a single conceptual map, linking concepts nodes in a graph. The categories emerged afterwards mostly from observing the following connectivity patterns in the graph: \emph{i) hubs:} a single concept is linked to many other concepts, as it is the case of ``teams'' in Figure~\ref{fig:concepts-people}, which is linked to another 7 concept nodes, with 13 references supporting such connections; and \emph{ii) cycles in the corresponding undirected graph}, such as the one present in Figure~\ref{fig:concepts-people}: ``(DevOps) $\rightarrow$ is a $\rightarrow$ (culture of collaboration) $\rightarrow$ shared across different types of $\rightarrow$ (teams) $\rightarrow$ may or may not have a $\rightarrow$ (DevOps role) $\rightarrow$ requires certain $\rightarrow$ (knowledge, skills, and capabilities for DevOps) $\leftarrow$ must teach $\leftarrow$ (programming education) $\leftarrow$ imposes challenges for $\leftarrow$ (DevOps).'' The presented hub and cycle examples led to the emergence of the \emph{people category}. We were driven by data to identify the following categories: \emph{process, people, delivery, and runtime}. We then fit other concepts into these categories. If some concepts did not fit in the found categories, new categories could be created, but this did not happen.

\end{versiontwo}

The process of building our conceptual maps also obeyed the following rules and restrictions:

\begin{itemize}
\item Nodes in the map contain \devops concepts.
\item The map can link nodes to describe relations among concepts.
\item A link can be read as a sentence binding the linked concepts, such as ``(DevOps)  $\rightarrow$ is a  $\rightarrow$ (culture of collaboration).''
\item Each core paper received a unique reference code.
\item Every concept and link must be supported by at least one core paper.
\item The link's label contains reference codes. 
\item A link can have different labels, each one with its references.
\item Some concepts, with common links, are grouped to save space in the figure.
\item A concept can belong to multiple categories. 

\end{itemize}

%The diagrams realizing our conceptual maps were produced using the software \texttt{Dia}.
%%TODO: comentado por Paulo

\begin{versiontwo}

Usually Grounded Theory procedures end with \emph{theoretical coding} leading to the formulation of a new theory by the development of hypotheses linking concepts and categories~\cite{stol2016grounded}. In this work, we do not aim at the formulation of a new DevOps theory nor to answer open questions posed by the literature. Instead, we use our conceptual framework on DevOps, built with open coding and constant comparison, as input to: \emph{i)} study and classify DevOps tools; \emph{ii)} identify practical DevOps implications and segment such implications for engineers, managers, and researchers; and \emph{iii)} discuss the main DevOps open challenges.

Moreover, by providing a global view of the DevOps field, we expect our conceptual maps to be used by engineers and managers to guide their continuing education, by teams to support reflection and continuous improvements of their DevOps journeys, and by academics to identify research topics, influences, and implications.

\end{versiontwo}

In the next section, we give an overview of articles analyzed to build our conceptual maps, as well as other relevant sources of knowledge on \devops. In Section \ref{sec:concepts}, we present the conceptual maps we produced to structure a set of relevant DevOps concepts grounded in the literature.

\section{Sources of knowledge}
\label{sec:sources}

A diverse community of academics and practitioners around \devops has been formed in recent years, and it is not very obvious where they publish their reports, research, and relevant information since there are very few \devops-specific venues. In this section, we present the sources of knowledge with the most impact, which include peer-reviewed papers, books, talks, and others.  We also categorize our core papers and provide an overview of the existing surveys of \devops, providing a list of related readings on the subject. We extract from this list the readings the \devops community consolidated as their references, which provides a reading guide for newcomers.  

\subsection{Peer-reviewed literature}

This survey primarily relies on evaluating peer-reviewed papers, which is standard practice in the academic community for guaranteeing the quality and credibility of published works.  

\begin{figure}[ht]
 \centering
  \includegraphics[scale=0.55]{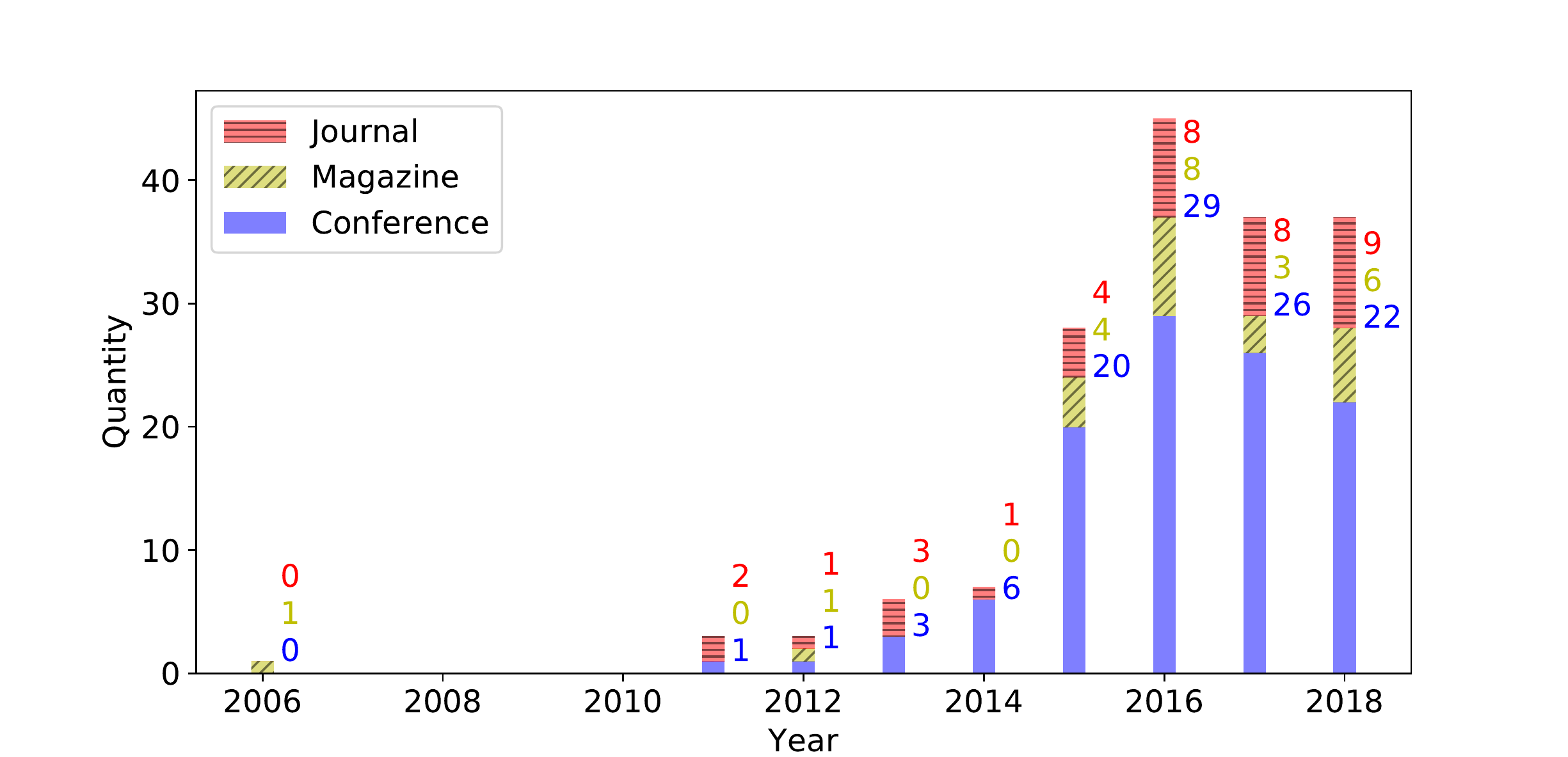}
  \caption{Publications by source types and publication year}
  \label{fig:source_types-per-year}
\end{figure}

Figure~\ref{fig:source_types-per-year} depicts the evolution of published works on \devops according to the criteria defined in our study design (papers collected in September 2018). Similar to most research areas, conferences generally publish preliminary results.  In contrast,  journals publish more mature studies. Academic journals primarily reach an academic audience, while magazines also have practitioners as their readers. 

The ``IEEE Software'' magazine was a constant source of papers on DevOps, having published 13 of the 50 core papers. They are mostly relatively short papers describing real experiences faced by organizations. Therefore, it is advisable for practitioners to subscribe to this source for the latest updates on DevOps.

We did not find any peer-reviewed journal, magazine, or event dedicated exclusively to the theme. The closest related events we found were the ``International Workshop on Release Engineering'' and the ``International Workshop on Quality-Aware DevOps.'' For researchers beginning a study on \devops, these are two potential venues for publishing their first results. The absence of a specific venue indicates an opportunity for the academic community to organize scientific events and journal special issues dedicated to this area.

After paper selection, following the protocol presented in Section \ref{sec:methodology}, we classified all \numberofmainselectedpapers papers into seven categories, as shown in Table~\ref{tab:papers-classification}. The categories are described in the sequence.

{
\small

\begin{table}[ht]
\centering
\caption{Classification of core papers}
\label{tab:papers-classification}
\begin{tabular}{l c r} \hline

\itshape{Category} & \itshape{Quantity} & \itshape{Papers} \\
\hline

% KEEP ORDERED
Practices & 15 & \citesel{wettinger2015knowledge, wettinger2015environments, woods2016view, basiri2016chaos, magoutis2015social, jaatun2018incident, rahman2018defective, li2018api, zhu2015frequency, brown2016microservices, cukier2013patterns, ebert2016devops, kang2016docker, kersten2018explosion, forsgren2018metrics} \\

Experience report & 8 & \citesel{balalaie2016microservices, callanan2016right, chen2015huge, feitelson2013facebook, gray2006conversation, neely2013easy, siqueira2018gov, cerqueira2015researchops} \\

Impact & 7 & \citesel{yasar2016security, shahin2016architecting, roche2013quality, jaatun2017security, rajkumar2016culture, sill2014standards, rossem2018virtualized} \\

Concepts & 7 & \citesel{humble2011devops, humble2017willwork, punjabi2016stories, bass2018architect, dyck2015release, debois2011revolution, pang2016maintenance} \\

Challenges & 6 & \citesel{laukkarinen2017medical, lwakatare2016towards, diel2016distributed, olsson2012stairway,leppanen2015highways, claps2015journey} \\

Education & 4 & \citesel{chung2016ksa, hussain2017zealand, bai2018feedback, christensen2016teaching} \\

Adoption & 3 & \citesel{nybom2016mixing, snyder2018analytics, feijter2018maturity} \\

\hline

\end{tabular}
\end{table}

% end small
}

\myparagraph{Practices} provides tools, practices, patterns, and strategies to enhance \devops. These studies offer guidelines for selecting deployment technology~\citesel{wettinger2015knowledge}, explanation of the use of metrics~\citesel{forsgren2018metrics}, how to support communication among global peers through specialized social networks~\citesel{magoutis2015social}, and how injecting infrastructure faults at production can support system reliability~\citesel{basiri2016chaos}.

\myparagraph{Experience Report} refers to reports of organizations adopting \devops or continuous delivery/deployment. For example,  Siqueira~\emph{et al.}  describe the impact of a \devops team in the context of a governmental project~\citesel{siqueira2018gov}.

\myparagraph{Impact} explores how \devops effects other aspects of software development such as architecture~\citesel{shahin2016architecting}, security~\citesel{yasar2016security}, quality assurance~\citesel{roche2013quality}, or the impact of \devops on software development in specific scenarios, such as the development of software for research~\citesel{cerqueira2015researchops}.

\myparagraph{Concepts} introduces basic concepts in \devops and continuous delivery, such as the deployment pipeline~\citesel{humble2011devops}.

\myparagraph{Challenges} covers the challenges entailed in adopting \devops or continuous delivery/deployment. Some papers focus on specific challenges, such as \devops in regulated domains~\citesel{laukkarinen2017medical}, embedded systems~\citesel{lwakatare2016towards}, or communication~\citesel{diel2016distributed}.

\myparagraph{Education} investigates the challenges in teaching \devops. Articles in this category propose teaching methods~\citesel{christensen2016teaching}, as well as explore the \textbf{K}nowledge, \textbf{S}kills, and \textbf{A}bilities (KSA) necessary for \devops professionals~\citesel{chung2016ksa,hussain2017zealand}.

\myparagraph{Adoption} covers the different models for \devops adoption~\citesel{nybom2016mixing} and describes how an organization chooses a set of metrics to evaluate its \devops adoption process~\citesel{snyder2018analytics}.  This subject is further discussed in Section~\ref{sec:devosp-adoption-approaches}.\\

\subsection{Related Surveys}
\label{sec:related-surveys}
Even though \devops is a recent research topic, several studies have performed a Systematic Literature Review (SLR) on the subject. They aim at finding a final \devops definition and which aspects influence its adoption on an organization. Table \ref{tab:research-questions} depicts the major research questions of these previous SLR. It is worth to mention most previous SLRs performed their literature review before 2015. It means they primarily examined the pioneering works on \devops and did not capture the significant growth of publications depicted in Figure~\ref{fig:source_types-per-year}; therefore, they miss most of the currently existing literature in the topic. This is evidenced by their conclusion that initial studies have a very low quality~\cite{Erich2017SLR}.

 {
\small

\begin{table}[ht]
\centering
\caption{Primary research questions from previous \devops SLR works}
\label{tab:research-questions}
 \begin{adjustbox}{max width=\textwidth}
\begin{tabular}{l r }
\hline
\emph{Research Question} & \emph{References}  \\ 
\hline

RQ1: What is the meaning of the term \devops? &  
\begin{tabular}{@{}c@{}}
\cite{travassos2016SLR, Erich2017SLR, gill2017SLR, Jabbari2016SLR, Bosch2017SLR} 
\end{tabular}  \\ \hline 

RQ2: What are the issues motivating the adoption of \devops?   &  
\begin{tabular}{@{}c@{}}\cite{travassos2016SLR, Erich2017SLR, gill2017SLR} \\  \end{tabular}  \\ 
\hline 

RQ3: What are the main expected benefits of adopting \devops?  &  
\begin{tabular}{@{}c@{}}\cite{travassos2016SLR, Erich2017SLR, gill2017SLR} \end{tabular}  \\ 
\hline 

RQ4: What are the main expected challenges/impediments of adopting \devops? & 
\begin{tabular}{@{}c@{}}\cite{Paasivaara2015LSR, travassos2016SLR, Erich2017SLR, gill2017SLR} \end{tabular}  \\ \hline 

%RQ5: What is known about the \devops Tools? & 
%\cite{gill2017SLR}  \\ 
%\hline 

%  \begin{tabular}{@{}c@{}} RQ6: What are the similarities and differences reported by %the authors of \\  primary studies between \devops and the other development methods? %\end{tabular} &
%\cite{Jabbari2016SLR} \\ 
%\hline

%\begin{tabular}{@{}c@{}} RQ7: What, if any, are the consensus definitions of %continuous practices and  \\  \devops, particularly in relation to one another, in %literature on con- \\ tinuous practices and \devops, and can such definitions be %found which \\ reflect mainstream usage yet cleanly separates the terms? %\end{tabular} &
%\cite{Bosch2017SLR} \\ 
%\hline

\end{tabular}
 \end{adjustbox}
\end{table}

% end small
}

The most relevant contribution of these previous studies is to correlate the absence of a well-established \devops definition with its effect on the perception of its benefits, expectations, and adoption challenges~\cite{Erich2017SLR}. 

To overcame the lack of the initial studies quality, they deployed methods other than Systematic Literature Review to uncover both academic and practitioners perspectives and to address the novelty of DevOps.
Table \ref{tab:reseach-sources} presents a list of sources used by these other works. Our survey selected works from all cited sources and had a much broader range of analyzed works than previous SLRs. We identified \numberofselectedpapers papers and carefully examined their titles and abstracts to select the \numberofmainselectedpapers core papers, which were then extensively analyzed to provide an updated overview of \devops.

{
\small

\begin{table}[ht]
\centering
\caption{Sources of \devops surveys}
\label{tab:reseach-sources}
 \begin{adjustbox}{max width=\textwidth}
\begin{tabular}{lcccccccc} \hline
\itshape{Author} & \itshape{Google Scholar} & \itshape{Springer Link} & \itshape{ACMDigital Library} & \itshape{IEEE Explore} & \itshape{Scopus} & \itshape{Web of Science} & \itshape{Others}                              & \itshape{Papers} \\ \hline
Fran\c{c}a \cite{travassos2016SLR}        & X             &               &                    &                                            &        &                &      X                               & 43    \\
\hline
Erich \cite{Erich2017SLR}         &                &               & X                  & X                                           & X      & X              &                                     &     27 \\
\hline
Smeds \cite{Paasivaara2015LSR}         & X              & X             & X                  & X                                           &        & X              & X & 27    \\
\hline
Ghantous \cite{gill2017SLR}         & X              & X             & X                  & X                                           &        &                & X                            & 30    \\
\hline
Stahl \cite{Bosch2017SLR}         &                &               &                    &                                               & X     &                &                                     & 35    \\
\hline
Jabbari \cite{Jabbari2016SLR}         &                &               & X                  & X                                           & X      &                &                                     & 49    \\
\hline
Lwakatare \cite{Lwakatare2015SLR}         & X              &               & X                  & X                                           & X      & X              & X                      & 22   \\
\hline
\textbf{This survey}         & X              & X              & X                  & X                                            & X      & X              &         X              & \textbf{\numberofselectedpapers (\numberofmainselectedpapers)} \\ \hline  
\end{tabular}
 \end{adjustbox}
\end{table}

% end small
}

Erich \textit{et al.} \cite{Erich2014SLR,Erich2017SLR} is the most referenced literature review of \devops. 
The authors performed an SLR~\cite{Erich2014SLR,Erich2017SLR} and conducted focused interviews on \devops principles and practices at six organizations~\cite{Erich2017SLR}.
From SLR, they extracted seven areas related to \devops: ``culture of collaboration, automation, measurement, sharing, services, quality assurance, and governance''~\cite{Erich2017SLR}. They conclude that there was, in general, low quality on the academic studies on \devops and a lack of studies to evidence DevOps' effectiveness. 
From focused interviews, they analyzed how organizations implemented \devops in four dimensions: governance, personal traits, department, and practitioners. They categorized answers from each organization into one of the seven areas extracted from SLR to portray their perceptions. 
Although both academic studies and organizations stated the benefits of adopting \devops, Erich \textit{et al.} concluded that no quantitative studies support such a claim~\cite{Erich2017SLR}, and reinforce the necessity of experimental studies to verify the effectiveness of \devops.

Fran\c{c}a et al.~\cite{travassos2016SLR} assumed academic studies were insufficient to address DevOps thoroughly, so
they alternatively conducted a Multivocal Literature Review (MLR), in which most of the sources were extracted
from gray literature, including books, websites, and industry journals and technical reports.
The authors describe \devops as a collection of principles, practices, required skills, and organizations' motivations to adopt it.
They identified seven areas related to \devops: social aspects, automation, measurements, sharing, quality assurance, and leanness.
Their main contribution was to describe DevOps characteristics associated with the practitioners' community and state-of-the-practice. 

\begin{versiontwo}
Erich \textit{et al.}~\cite{Erich2017SLR} and Fran\c{c}a et al.~\cite{travassos2016SLR} indicate the need for more research on \devops, especially quantitative research to evaluate DevOps' effectiveness. Even with distinct methodologies, these two works arrived at similar conclusions. They concentrate their studies on the \textit{Process} and \textit{People} categories of our conceptual map, to be described in Section \ref{sec:concepts}. Previous surveys~\cite{gill2017SLR,Jabbari2016SLR,travassos2016SLR} did not cover the \textit{Runtime} and \textit{Delivery} categories of our conceptual map; they have only mentioned some of the concepts. None of the previous SLRs have discussed the technical implications and complexity of adopting \devops practices such as automation, microservices architectures, containerization, and toolset management. The concepts and implications of practical aspects of \devops not only empower managers to make assertive and strategic decisions but also equip engineers with best practices for \devops adoption. We provide lessons learned of \emph{i)} using \devops with legacy systems; \emph{ii)} the complexities and mistakes when adopting the microservice architecture; \emph{iii)} guidelines for automating the delivery pipeline; \emph{iv)} the advantages and drawbacks of teams autonomy in toolset selection.
These points form the major contribution of our survey, wherein we deepen the analysis of practical aspects of \devops adoption with the aid of a conceptual map of \devops to guide engineers, managers, and researchers toward mastering skills necessary to enable \devops. Additionally, the quality of studies on \devops improved in recent years, so we have updated the implications of deploying \devops in an organization as well as contributed to more technical implications for practitioners and researchers, such as re-designing systems architecture, deployment pipeline, and quantitative assessment metrics. Previous SLRs did not cover these latest subjects.

\end{versiontwo}

\subsection{Books}

\begin{figure}[ht]
  \includegraphics[scale=0.5]{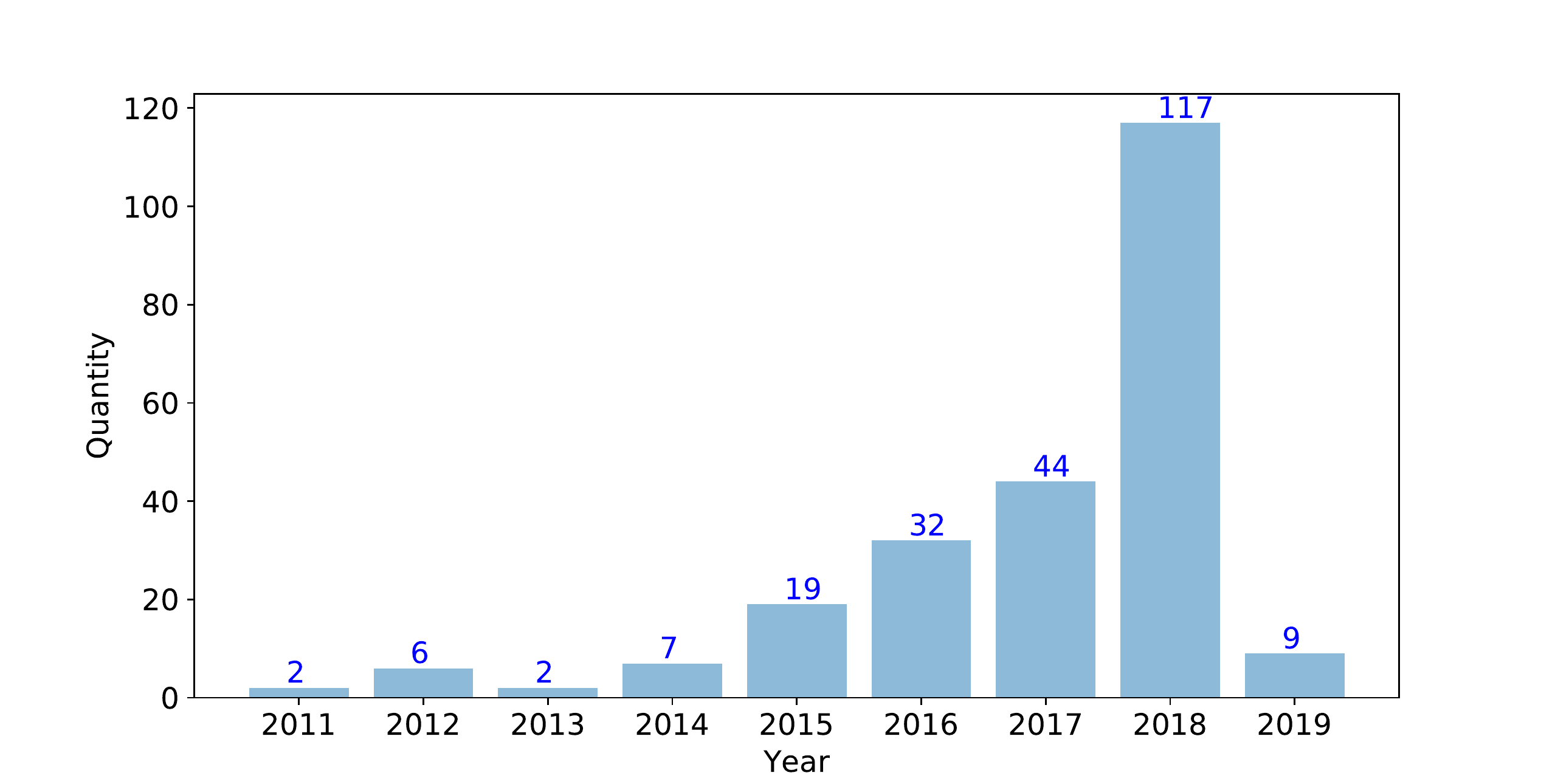}
  \caption{Distribution of books about DevOps by publication year according to Amazon.com}
  \label{fig:books-per-year}
\end{figure}

Books are the standard source of knowledge on \devops among practitioners. Although they usually have more pages than academic papers,  they are easier to read and to assimilate. We searched Amazon.com on December 2018, for books written in English containing ``\devops'' in the title. Figure~\ref{fig:books-per-year} presents the annual distribution of a total of 238 books found in this search. Comparing to Figure~\ref{fig:source_types-per-year}, it is possible to see their shapes roughly shift by one year, showing how academic publication precedes book publication.

One of the seminal books on \devops is ``Continuous Delivery''~\cite{humble2010continuous}, from 2010. It does not explicitly use the word ``\devops,'' but it describes in detail the deployment pipeline pattern, which is usually central to \devops strategies.
The oldest book found at Amazon containing the word ``\devops'' in its title is ``DevOps: High-impact Strategies''~\cite{roebuck2011devops} from 2011.
Some of the books frequently cited by our reviewed literature are ``The Phoenix Project''~\cite{Kim2014Phoenix} and ``DevOps: A Software Architect's Perspective''~\cite{Bass2015Devops}.
The most popular books at Amazon containing the word ``\devops'' in the title are ``Accelerate: The Science of Lean Software and DevOps''~\cite{forsgren2018accelerate}, ``The Phoenix Project''~\cite{Kim2014Phoenix}, and ``The DevOps Handbook''~\cite{kim2016handbook}.

It is worthy to note that three authors of these cited books (Nicole Forsgren, Jez Humble, and Gene Kim) also collaborate with the State of \devops Reports~\cite{puppet2014devops,puppet2016devops}, a vital source of knowledge on \devops, as discussed next.

\subsection{Other sources}

\myparagraph{State of DevOps Reports} annual industrial research published since 2013 by renowned authors in the DevOps field based on the survey of multiple organizations worldwide~\cite{puppet2014devops, puppet2016devops}. The reports provide proper usage of quantitative metrics for assessing \devops initiatives. More about this will be discussed in Section~\ref{sec:assessing-devops}.

\myparagraph{The Twelve-Factor App} a manifest with architectural principles aligned with the \devops philosophy~\cite{wiggins2011twelve}. It is a reputable source amongst practitioners and is typically used to judge architectural decisions.  The present instructions like: (a) keep development, staging, and production environments as similar as possible, and (b) use environment variables to store configuration that varies across environments, among others. 
A related set of principles is the Reactive Manifesto, which is also used by practitioners as a guideline~\cite{boner2014reactive}. 

\myparagraph{Talks and Videos} there are many practitioners' conferences about DevOps and conferences that include DevOps talks. Many practitioners rely on these talks for up-to-date information on DevOps, especially when the speakers are from reputable companies, such as Amazon~\cite{amazon2015devops}, Microsoft~\cite{microsoft2018devops}, Google~\cite{google2017devops}, Netflix~\cite{netflix2018devops}, and Opscode~\cite{chef2015kungfu}. Many of these conference talks are made available on the web,  where they join many other short videos about DevOps. On Youtube or other video platforms, dozens of videos have titles matching the ``What is DevOps'' string. 

\myparagraph{Global community} engineers leverage global-community knowledge to overcome their daily challenges. This knowledge comes from code repositories, such as \texttt{GitHub} and \texttt{Chef Supermarket}, and from discussion forums like \url{stackoverflow.com}~\citesel{magoutis2015social}. In this way, engineers interact not only with colleagues within the enterprise but also with peers around the world, building a global community.

\section{Fundamental concepts}
\label{sec:concepts}

\vtwo{In this section, we present a \emph{conceptual framework} of fundamental concepts of \devops.} The concepts emerged from our systematic analysis of the literature, which followed the protocol presented in Section \ref{sec:methodology}. 
Our contribution is to provide a relevant and structured set of concepts on \devops, grounded in the literature, to support analysis and understanding of \devops challenges.

\vtwo{Our conceptual framework~\cite{miles1994qualitative} on DevOps is composed of a conceptual map outlining the conceptual categories and of other four conceptual maps that are diagrams structured as graphs in which nodes depict concepts and arrows represent relationships among concepts.} There is always at least one core bibliographic reference supporting a particular relationship. The references are represented by codes that can be identified in the List of Core Papers on page \pageofmainselecetdstudies.

The concepts are distributed into four major categories: process, people, delivery, and runtime. The \emph{process} category encompasses business-related concepts. \emph{People} covers skills and concepts regarding the culture of collaboration. \emph{Delivery} provides the concepts necessary for Continuous Delivery, and, finally, \emph{runtime} synthesizes concepts necessary to guarantee the stability and reliability of services in a continuous delivery environment. 

\begin{figure}[ht]
  \includegraphics[scale=0.4]{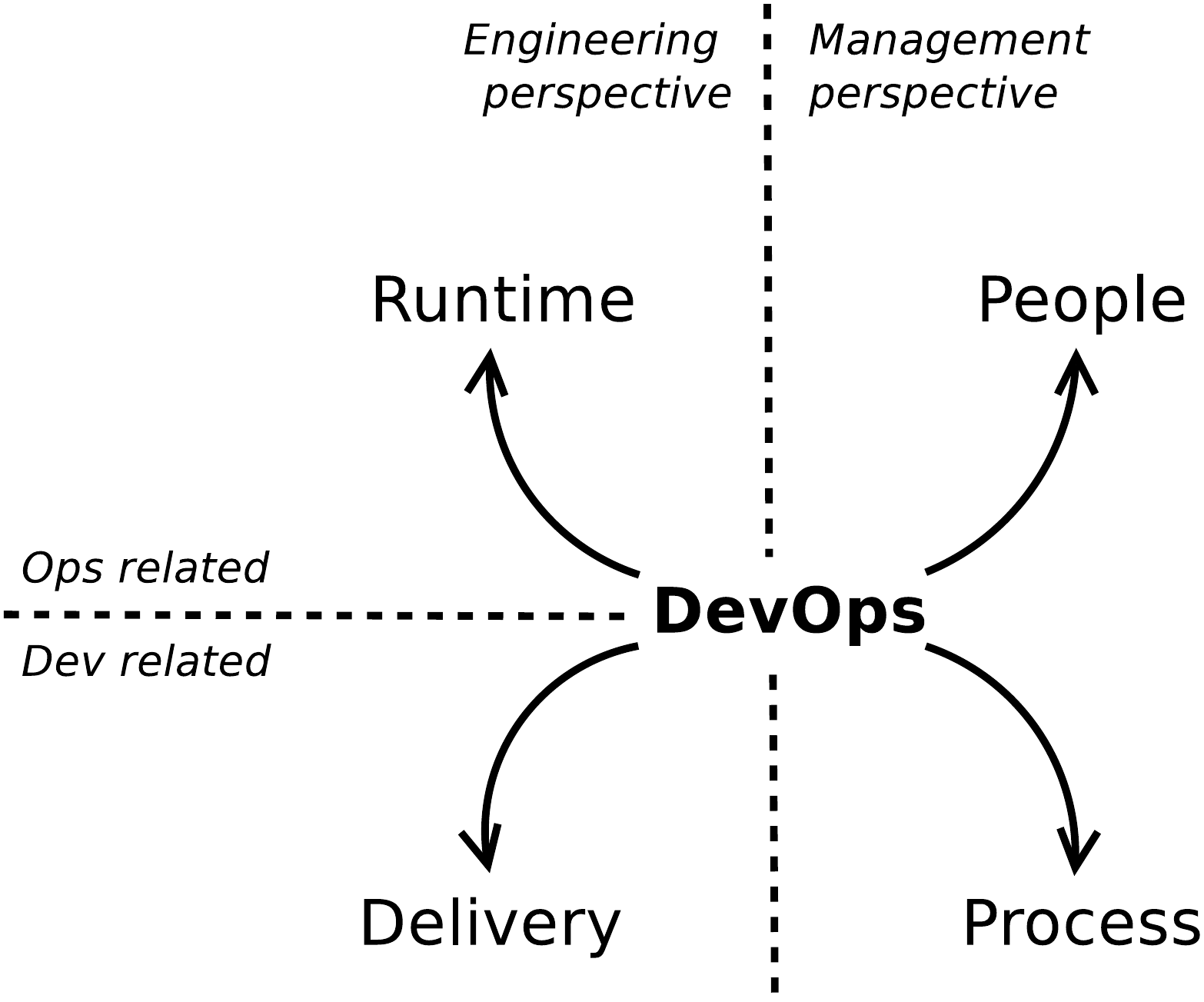}
  \caption{\devops overall conceptual map}
  \label{fig:concepts-families}
\end{figure}

While the process and people categories relate more to the management perspective, runtime and delivery relate more to an engineering perspective. Moreover, while \emph{delivery} concepts relate more to developers, \emph{runtime} concepts relate more to the traditional operator role. The categories are depicted in Figure~\ref{fig:concepts-families} and described in this section.

\begin{versiontwo}
Before proceeding to the categories descriptions, we briefly present some reasons to support that these categories are enough to frame the DevOps field: \emph{i)} it seems reasonable to split ``dev'' and ``ops'' concepts as induced by the categories \emph{delivery} and \emph{runtime}; \emph{ii)} it makes sense to separate more technical concepts from less technical concepts, as caused by the separation between the perspectives of engineering and management. \emph{iii)} These categories match our DevOps definition: ``a collaborative and multidisciplinary effort within an organization'' regards \emph{people}; ``continuous delivery'' is a \emph{process} that is ``automated'' by the \emph{delivery} techniques, which also guarantee software ``correctness''; finally, software ``reliability'' is promoted by the \emph{runtime} concepts.
\end{versiontwo}

\subsection{Process}

\begin{figure}[ht]
  \includegraphics[scale=0.25]{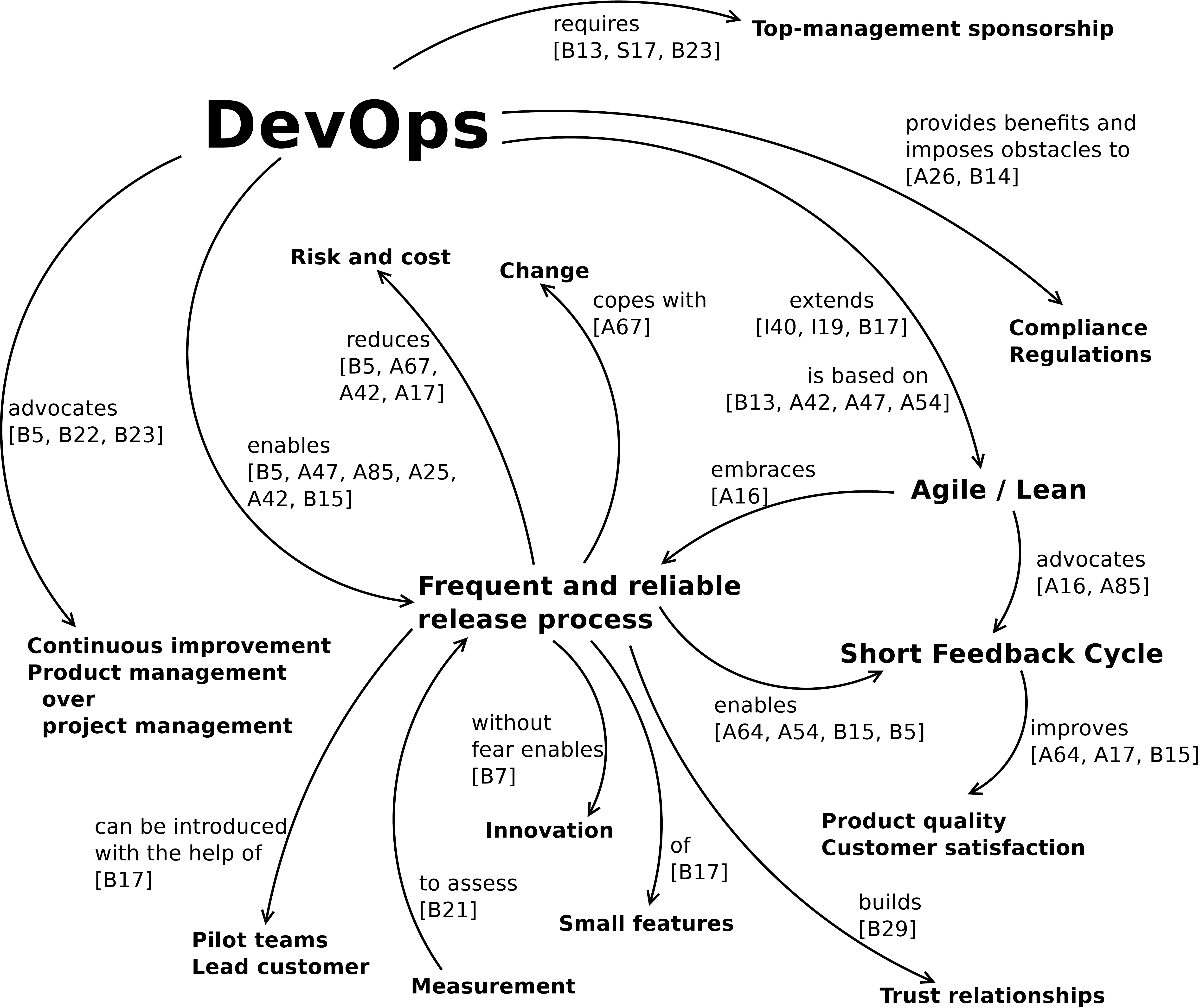}
  \caption{Process conceptual map}
  \label{fig:concepts-process}
\end{figure}

\devops aims to achieve some business outcomes, such as reducing risk and cost, complying with regulations, and improving product quality and customer satisfaction. We grouped these concepts into the \emph{process} category of concepts, presented in Figure~\ref{fig:concepts-process}, which reveals that \devops achieves such business outcomes through the accomplishment of a process with frequent and reliable releases. In particular, the diagram explicit that continuous delivery leads to product quality and customer satisfaction because of the short feedback cycle it provides.

One could argue that rigorous human and hierarchical approval processes can reduce risk, comply with regulations, and provide product quality. Nevertheless, \devops is different, once its practices are based on agile and lean principles, which embrace change and shorten the feedback cycle, as depicted in the diagram. 

\subsection{People}

\begin{figure}[ht]
  \includegraphics[scale=0.25]{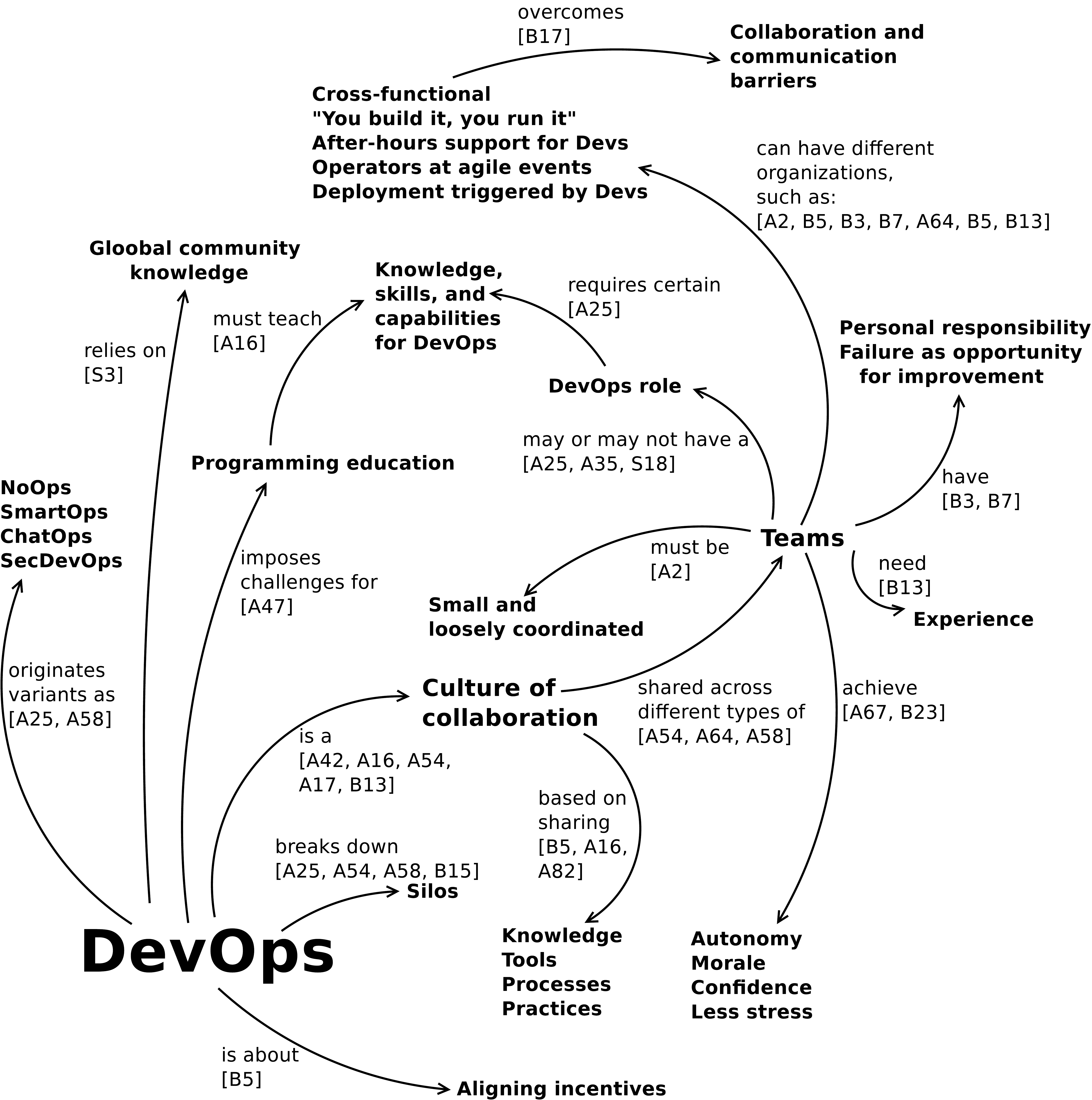}
  \caption{People conceptual map}
  \label{fig:concepts-people}
\end{figure}

The term ``\devops'' centers on the idea of bringing together \emph{people} from development and operations through a culture of collaboration. The concepts around this idea were grouped in the \emph{people} category, which is presented in Figure~\ref{fig:concepts-people}.  \vtwo{It depicts that \devops intends to break down the walls among silos, aligning the incentives across the organization.} However, breaking down silos raises many questions concerning concepts depicted in the diagram: how can organizations perform a cultural shift that implies more responsibilities for developers? How do developers acquire operations skills? Can developers and operators work in the same team and still keep different job titles? Should ``\devops'' be a role? Should teams be cross-functional and include operators? Should an operator be exclusive to one team? What does it mean being an operator in the \devops context? Although the concepts of this category are frequent in the literature, the answers to the previous questions are not clear yet, so we debate them in Section~\ref{sec:devosp-adoption-approaches}. 

\subsection{Delivery}

\begin{figure}[ht]
  \includegraphics[scale=0.25]{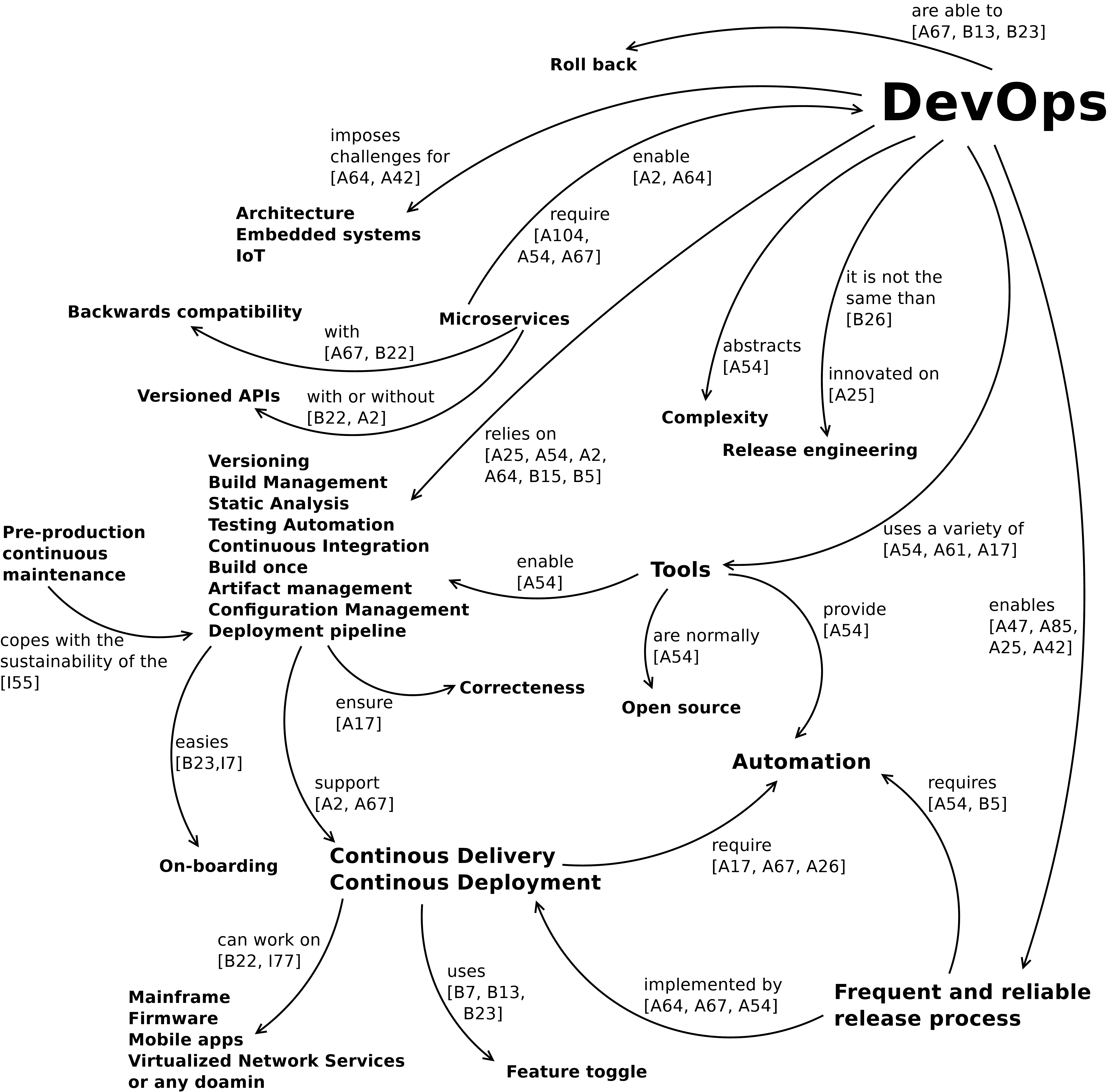}
  \caption{Delivery conceptual map}
  \label{fig:concepts-delivery}
\end{figure}

The core strategy for achieving a frequent and reliable delivery process is the automation of the \emph{deployment pipeline}, from which \devops tools and techniques emerge. We grouped the concepts surrounding the deployment pipeline into the \emph{delivery} category, which is presented in Figure~\ref{fig:concepts-delivery}.  Automation tools, as depicted in the diagram, are usually open source and enable core \devops practices, such as versioning, testing automation, continuous integration, and configuration management. The diagram also shows the interconnection between \devops and microservices: one supports the other, as we explore in Section~\ref{sec:redesign}. Other concepts related to microservices are also depicted, such as backward compatibility and API versioning.

There is still some debate on concrete strategies for tooling, such as the containerized approach leveraged by Docker on the one hand, and continuous configuration convergence, such as in Chef and Puppet, on the other. We discuss such debates in Section~\ref{sec:tools}. However, we claim the \emph{delivery} concepts are much more stable and accepted by the community than the \emph{people} concepts. 

\subsection{Runtime}

\begin{figure}[ht]
  \includegraphics[scale=0.25]{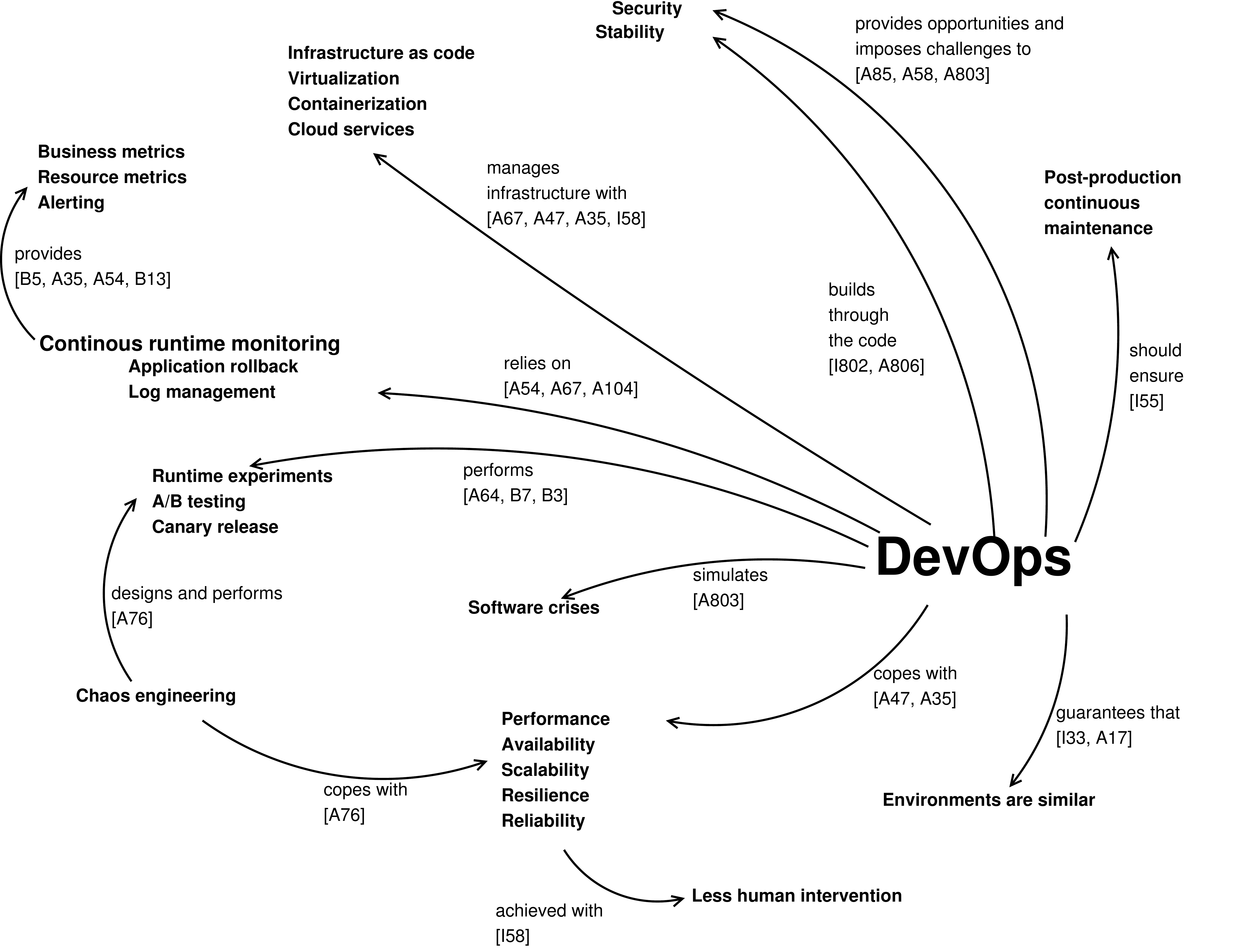}
  \caption{Runtime category of concepts}
  \label{fig:concepts-runtime}
\end{figure}

It is not enough to continuously deliver new versions, but it is also necessary for each new version to be stable and reliable. Thus, \emph{runtime} concepts are a subsequent and necessary extension of \devops. The \emph{Runtime} category of concepts is presented in Figure~\ref{fig:concepts-runtime}, which shows the desired outcomes, such as performance, availability, scalability, resilience, and reliability. The figure also shows paths to achieve the mentioned outcomes, like the use of infrastructure-as-code, virtualization, containerization, cloud services, and monitoring. As depicted by the diagram, \devops can monitor high-level business metrics or low-level resource metrics. Another topic, also shown in the figure, is to run experiments in the production environment, like injecting failures to ensure software reliability, as advocated by the chaos engineering approach~\citesel{basiri2016chaos}. All of this reduces human intervention to ensure software reliability, which is another factor that challenges the traditional role of operators, as we argue in Section~\ref{sec:devosp-adoption-approaches}. 

\subsection*{}

We complement our conceptual analysis with the practical side of \devops by investigating \devops' tools and actors, and their relations to concepts. As the presentation of \devops' concepts has already raised some concerns that demand further discussion, some issues will also emerge from the discussion on DevOps tools in the next section.

\section{Toolset}
\label{sec:tools}

We define ``DevOps tools'' as the tools pursuing one of the following goals: 1) assisting \emph{human collaboration} across different departments; 2) enabling \emph{continuous delivery}; or 3) maintaining software \emph{reliability}. Our contribution is to associate tools to concepts presented in Section \ref{sec:concepts}. This correlation between tools and concepts equips and empowers practitioners for well-informed decisions on tooling, and thereby enforces alignment between technical, managerial and organizational decisions. This conceptual approach becomes immensely important when considering the ever-changing nature of \devops tools~\citesel{shahin2016architecting}. 

Although tools matter, caution is required to avoid turning them into the core of a \devops strategy. Adopting new tools demands significant effort, and applying this effort without considering the expected outcome can lead to considerable waste. For example, a \devops adoption process must consider team structuring as more critical than the choice of specific tools.

\devops heavily relies on automation tools. The choice of a proper toolset imposes challenges for \devops professionals. It is not feasible for a single person, even an organization, to master all the \devops tools, given the vast number of available tools~\citesel{ebert2016devops,kersten2018explosion} and their rapid evolution~\citesel{shahin2016architecting}. 

We also contribute to discussing the most important categories of \devops tools, presented in Table~\ref{tab:tools}, \vthree{by adding the following complementary information to each category: \emph{i)} some of the most used tools in each category; \emph{ii)} the intended users: whether developers or operators; \emph{iii)} the goals: \emph{human collaboration}, \emph{continuous delivery} or \emph{reliability}; and \emph{iv)} \devops concepts.}

% by adding some complementary information to each category:

% \begin{itemize}
% \item Some of the most used tools in each category.
% \item The intended users: whether developers or operators.
% \item The goals: \emph{human collaboration}, \emph{continuous delivery} or \emph{reliability}.
% \item \devops concepts.
% \end{itemize}

The complementary information, like tool examples, helps the reader to understand each category, leading to the following additional contributions: \emph{i)} helping \devops practitioners focus on organization requirements, while choosing the most suitable set of tools; \emph{ii)} helping the reader to grasp a critical view about the objective of each tool and how to use it in an organization; \emph{iii)} showing how the frontiers between developers and operators have been blurred in the \devops context through toolset usage. 

\subsection{Tools for knowledge sharing}

The \devops strategy focuses on human collaboration across silos. Concretely, this entails different departments sharing knowledge, process, and practices, which requires sharing specific tools. 

We cite \texttt{GitLab} and \texttt{Rocket Chat} as examples of tools in this category. Gitlab not only provides a code repository but also has a wiki system that enables developers and operators to share knowledge. Gitlab offers an issue management system, which is typically used by developers and business analysts to share knowledge through agile practices like issues comments and pull requests processes. Making operators also familiar with these issues helps them to understand the product and project contexts. Moreover, documenting production problems known by operators in the issue management system is an essential step toward handling non-functional requirements prioritization in the development flow. ChatOps (Chat and Operations) is a model that connects people,  tools,  process, and automation through conversation-driven interactions mediated by tools~\cite{mulyana2018chatops}. Rocket Chat is a communication tool that implements ChatOps by enabling a higher degree of automation and tool integration through the use of webhooks and chatbots.

These tools support \emph{human collaboration} and the \emph{people} category of concepts. However, despite the centrality of human collaboration in \devops strategy, the ``\devops Periodic Table''~\cite{xebia2016table}, a well-known list of more than 100 \devops tools, includes only eight tools for collaboration. This fact may suggest that human collaboration and \emph{people} concepts have been eclipsed in the \devops movement by the more technical goals and concepts.

\subsection{Tools for source code management}

Source code management tools usually intend to promote collaboration among developers. These tools are basic blocks to implement continuous integration and, therefore, \emph{continuous delivery}. From this traditional perspective, one could relate source code management solely to the \emph{delivery} category of concepts.

However, source code management can also be used by operators to store artifacts and automation scripts and to access software information that can impact operations activities. For example, a development bug may cause a memory leak that impacts operations. When storing infrastructure related artifacts, whether in the infrastructure-as-code style or just as plain text, operators provide developers better insight into how the software is executed. Therefore, source code sharing among developers and operators becomes a real point of collaboration.

The most traditional tools in this category are \texttt{SVN} and \texttt{Git}. More complete platforms, such as \texttt{GitLab} and \texttt{GitHub}, can also wrap Git, providing easier-to-use visualizations for code changes, as well as integrating with additional tools like the issue manager.

\subsection{Tools for the build process}

Build tools are highly developer-centered. Their goals relate to enabling \emph{continuous delivery} and the \emph{delivery} category of concepts. We have considered not only tools that generate deployable packages, but also called \emph{builds}, but also tools that generate essential artifacts and feedback using the source code as input.

Each programming language has some tools to cope with the build process by supporting dependencies resolution, implementation of custom tasks, or generation of deployable packages. Examples of dependency managers, also called package managers, are \texttt{Pip} for Python, \texttt{RubyGems} for Ruby, \texttt{NuGet} for .NET, and \texttt{Maven} and \texttt{Gradle} that compete in the Java landscape. Gradle eases the implementation of custom tasks during the build, as also done by GNU Make, that is often used for GNU/Linux packages, or Rake for Ruby. Some of the cited tools, such as Maven, are also responsible for producing the deployable package, such as WAR files for Java environments. However, for some languages, like Python and Ruby, there is no need for producing a deployable package as a single-file artifact. It is also possible to use more generic package tools coupled to the target environment, such as \texttt{Debian} packages. 

Each programming language also has some unit-test frameworks, which provide vital feedback for developers about the correctness of the software. Some examples are \texttt{JUnit} and \texttt{Mockito} for Java, \texttt{RSpec} for Ruby, and \texttt{NUnit} for .NET. More sophisticated testing which automates the end-user behavior for web applications is possible with browsing automation tools, such as \texttt{Selenium}.

Another type of feedback for developers is regarding source code quality, which is provided by source-code static analysis tools, such as \texttt{SonarQube} and Analizo~\cite{terceiro2010analizo}. SonarQube classifies code problems and evaluates coverage metrics as well as technical debts in several programming languages via its plugins. Analizo supports the extraction of a variety of source code metrics for C, C++, C\#, and Java codes. In general,  source-code analysis tools can point to issues in source-code, like non-compliance to the standard style, problems in maintainability, risk of bugs, and even vulnerability breaches. Source-code static analysis tools vary in supported programming languages and in the way they are delivered. They can be provided as-a-service, for example, \texttt{Code Climate}, or even within the developer environment through tools such as \texttt{PMD} for the Eclipse IDE.

One critical requirement for a build tool in \devops context is automation. There are some products whose build actions can be only triggered through a Graphical User Interface (GUI), which is not acceptable in a continuous delivery process. 

\subsection{Tools for continuous integration}

Continuous integration tools orchestrate several automated actions that, together, implement the \emph{deployment pipeline} pattern. Among the stages orchestrated by the pipeline are: package generation, automated test execution for correctness verification, and deployment to both development and production environments. These tools are related to the \emph{delivery} category of concepts. 

The main actors responsible for defining the pipeline structure are typically developers. Operators usually collaborate on defining the deployment stages; they are also in charge of maintaining the continuous integration infrastructure running as a service to developers. For this, operators run continuous integration tools such as \texttt{Jenkins} or GitLab CI (\url{about.gitlab.com/features/gitlab-ci-cd}). Deployable packages can be stored on repositories like \texttt{Nexus} for enabling future rollback.

Since downtime of continuous-integration infrastructure results in the interruption of continuous delivery, it is common to use highly-available third-party services, such as GitLab.com and \texttt{Travis}. 

\subsection{Tools for deployment automation}

Deployment automation tools are employed in the deployment stages of the pipeline to make the \emph{continuous delivery} process viable. They enable frequent and reliable deployment processes, as well as other concepts related to the \emph{delivery} category.  Although the use of deployment automation tools is a joint effort between developers and operators, the primary mission of continuous delivery is putting the deployment schedule under business control.  

Every automated deployment approach relies on the concept of ``infrastructure as a code.'' It requires engineers to specify servers, networking, and other infrastructure elements in files modeled after software code~\cite{morris2016infraascode}. In this model, deployment and infrastructure definitions are shared among developers and operators, allowing effective cooperation between them.

Automated deployment can encompass not only the deployment of the package to the production environment but also the provisioning of the target environment. Such provisioning is usually performed in cloud environments~\citesel{cukier2013patterns}, such as \texttt{Amazon Web Services}, \texttt{Google Cloud}, and \texttt{Azure}. These platforms deliver a vast amount of infrastructure services via Infrastructure as a Service (IaaS) model, like launching virtual machines, databases, queues, etc. The creation of all these resources can also be orchestrated, in Amazon's platform, through the usage of \texttt{AWS Cloud Formation}. Operators and developers can share the task of using IaaS services.

It is also possible to use a Platform as a Service (PaaS), such as \texttt{Heroku}. In the PaaS approach, the platform is responsible for the deployment, and developers do not know the underlying virtual infrastructure.  
Serverless services -- a new trend in cloud computing -- aim to abstract servers for developers, which is close to PaaS; the main differences are the scaling transparency and the charge for computing usage~\cite{roberts2018serverless}. These models suggest there is no need for specialized operations teams~\citesel{cukier2013patterns}; or, at least, that fewer people are required for operations~\cite{roberts2018serverless}. 

If using IaaS, the environment provisioning is followed by the execution of scripts that effectively install the application in the target environment. A deployment script can be written using Shell Script, but configuration management tools such as \texttt{Chef} and \texttt{Puppet} offer advantages by leveraging operation system portability and idempotence mechanisms, which are difficult to achieve with Shell Script. An example of language construct of Chef that leads to portability is using the ``package'' resource, which is resolved to a concrete package manager, such as \texttt{Apt}, only at execution time. An example of an idempotent mechanism of Puppet in the ``service'' resource is declaring the desired final state of the service as ``running'' rather than writing a command to start the service~\cite{sato2014devops}. 

Another alternative for deployment is containerization, primarily implemented by \texttt{Docker}. Docker containers resemble virtual machines. The main difference is that they are more lightweight~\citesel{kang2016docker}, considering they share the kernel of the host. Docker and related tools, such as \texttt{Docker Compose}, \texttt{Kubernetes}, and \texttt{Rancher}, allow the specification of containers and their dependency relationships. These specifications generate container images in the build stage, which are later instantiated in the target environment. Docker has been used not only for deploying applications but also for deploying the underlying infrastructure~\citesel{kang2016docker}.

Containerization could be seen as complementary to the deployment script strategy (done by Chef or Puppet). However, in practice, these strategies seem to compete. A Chef script is executed continuously on the target node, and its success depends on the previous target node state. On the other hand, in containerization the whole containerized environment is generated in the build, so the environment is destroyed and rebuilt at each new software version. When compared to the Chef strategy, Docker yields faster and more reliable deployment~\citesel{kang2016docker}, but at the expense of bigger builds. 

Complimentary usage of configuration management and containers entails the setup of Docker environment and update of the operating system software, like SSH. Another usage is to manage configuration that varies across diverse environments (testing, staging, production). Such a configuration cannot be embedded in container images since the same image must be deployed in every environment. Nonetheless, the debate about configuration management versus containers is not fully explored in the literature. Among practitioners, although it is possible to find support for a complementary usage of both~\cite{barua2015role}, there is a trend to favor containers over configuration management tools~\cite{dougherty2015chefvsdocker,horowitz2018antipattern,kelly2016better}.

Zhu~\emph{et al.} compare the provisioning reliability of two approaches: using ``lightly baked images'' versus ``heavily baked images''~\cite{zhu2015frequency}. In the lightly baked approach, a virtual machine image is instantiated, and Chef is executed to install the application, whereas in the heavily baked approach the image already contains the entire application. This heavily baked approach is similar to the container strategy, but using virtual machine technology. The authors conclude that deployment using lightly baked images is less reliable since it involves more external resources during deployment, which tends to increase errors and delays.

As the application evolves, the database structure evolves as well. Traditionally,  ``database administrators'' maintain the database structure. The adoption of emergent design~\cite{martin2008emergence} and the need for frequent releases have encouraged the use of database migrations tools such as \texttt{Flyway}, which controls the automated application of schema updates across different environments, thus enabling developers to manage database structure themselves. In this context, one must rethink the traditional role of database administrators.

Developers and operators share these tools for deployment.  The main risk is the absence of clarity of responsibilities, which can cause friction in organizations during \devops adoption~\citesel{nybom2016mixing}. Developers believe they do not control their application and feel they are doing someone else's job. Additionally,  operators may judge developers as unable to perform some tasks or operate some tools.

\subsection{Tools for monitoring}

\emph{Monitoring} tools usually track applications' non-functional properties, such as performance, availability, scalability, resilience, and \emph{reliability}. Self healing,  alerts, log management, and business metric follow up are example tasks performed by monitoring tools; they relate to the \emph{runtime} category of concepts. 

Example of tools for monitoring and alerting are \texttt{Nagios}, \texttt{Zabbix}, and \texttt{Prometheus}. Examples of log management tools are \texttt{Graylog} and \texttt{Logstash}. Cloud services also play an essential role in guarantying non-functional properties of applications since they provide elastic resources that can be allocated on demand~\cite{nist2011cloud}. It is also typical for cloud services to provide monitoring and alerting services.

The tendency toward cross-functional, full stack teams~\cite{balalaie2016microservices,ebert2016devops}, combined with the expectation that developers must be accountable for the product~\citesel{gray2006conversation,feitelson2013facebook} pushes the use of these tools to the development team. Therefore, once again, the responsibility for using and mastering specific tools is unclear.

{
\small

\begin{table}[ht]
\centering
\caption{Major \devops tools related to actors and \devops concepts}
\label{tab:tools}
\begin{tabular}{l l l l l} \hline
\emph{Category} & \emph{Examples} & \emph{Actors} & \emph{Goals} & \emph{Concepts} \\ \hline

\begin{tabular}{@{}l@{}}Knowledge \\ sharing\end{tabular} &
\begin{tabular}{@{}l@{}}Rocket Chat\\ GitLab wiki \\ Redmine \\ Trello \end{tabular} &  
\begin{tabular}{@{}l@{}}Everyone \end{tabular} & 
\begin{tabular}{@{}l@{}}Human \\ collaboration\end{tabular} &
\begin{tabular}{@{}l@{}}Culture of collaboration\\ Sharing knowledge\\ Breaking down silos\\ Collaborate across departments\end{tabular}\\
\hline 

\begin{tabular}{@{}l@{}}Source code \\ management\end{tabular} &
\begin{tabular}{@{}l@{}}Git\\ SVN \\ CVS \\ ClearCase \end{tabular} 
&  Dev / Ops &
\begin{tabular}{@{}l@{}}Human \\ collaboration \\  \\ Continuous \\ delivery\end{tabular} &
\begin{tabular}{@{}l@{}}Versioning\\ Culture of collaboration\\ Sharing knowledge\\ Breaking down silos\\ Collaborate across departments\end{tabular}\\
\hline 

\begin{tabular}{@{}l@{}}Build \\ process\end{tabular}  &
\begin{tabular}{@{}l@{}}Maven\\ Gradle\\ Rake\\ JUnit\\ Sonar\end{tabular} 
&  Dev &
\begin{tabular}{@{}l@{}}Continuous \\ delivery\end{tabular} &
\begin{tabular}{@{}l@{}}Release engineering\\ Continuous delivery\\ Automation\\ Testing automation, Correctness\\ Static analysis\end{tabular} \\
\hline 

\begin{tabular}{@{}l@{}}Continuous \\ Integration\end{tabular} &
\begin{tabular}{@{}l@{}}Jenkins\\ GitLab CI\\ Travis\\ Nexus\end{tabular}   
& Dev / Ops &
\begin{tabular}{@{}l@{}}Continuous \\ delivery\end{tabular} &
\begin{tabular}{@{}l@{}}Frequent and reliable release process\\ Release engineering\\ Continuous integration\\ Deployment pipeline\\ Continuous delivery, Automation\\ Artifact management \end{tabular}   \\
\hline 

\begin{tabular}{@{}l@{}}Deployment \\ automation\end{tabular} &
\begin{tabular}{@{}l@{}}Chef, Puppet\\ Docker \\ Heroku \\ Open Stack \\ AWS Cloud Formation\\ Rancher \\Flyway\end{tabular} 
&  Dev / Ops &
\begin{tabular}{@{}l@{}}Continuous \\ delivery \\  \\ Reliability\end{tabular} &
\begin{tabular}{@{}l@{}}Frequent and reliable release process\\ Release engineering\\Configuration management\\ Continuous delivery\\ Infrastructure as code\\ Virtualization, Containerization\\ Cloud services, Automation\end{tabular} \\
\hline 

\begin{tabular}{@{}l@{}}Monitoring \\ \& Logging\end{tabular} &
\begin{tabular}{@{}l@{}}Nagios\\ Zabbix\\ Prometheus \\ Logstash\\ Graylog\end{tabular}  
& Ops / Dev &
Reliability &
\begin{tabular}{@{}l@{}}You built it, you run it\\ After-hours support for Devs\\ Continuous runtime monitoring\\ Performance, Availability, Scalability\\ Resilience, Reliability, Automation\\ Metrics, Alerting, Experiments\\ Log management, Security \end{tabular} \\
\hline 

\end{tabular}
\end{table}

% end small
}

\subsection{Actors}

The still undefined division of responsibilities in the DevOps world makes it difficult to associate a role to each DevOps tool. At the same time, the variety of DevOps tools seems to challenge the idea of a single person holding the title of a ``DevOps engineer.'' Even a whole cross-functional team may struggle to know all these tools. On the other hand, companies preserving the operations department may have difficulty in precisely defining which roles should use which tools.

Humble and Molesky suggest operations teams should provide IaaS to product teams~\citesel{humble2011devops}, such as continuous integration platform, compute provisioning, and monitoring services. Traditional product teams become responsible for using these tools, whereas operations teams are responsible for set-up and maintenance, ensuring their performance and availability, and consulting in their usage. In this scenario, the operations team becomes itself a product team~\citesel{humble2011devops}.

This pattern nears that of adopting outsourced development infrastructures, such as GitHub or GitLab. However, outsourcing the infrastructure has additional benefits, like freeing the team to experiment with different tools~\citesel{gray2006conversation}, and not only the ones supported by the operations staff. Most of the time, reputable infrastructure providers have better availability than in-house solutions. We discuss further the possible \devops adoption approaches and their consequences in Section~\ref{sec:devosp-adoption-approaches}.

\section{Implications for engineers, managers, and researchers}
\label{sec:implications}

\begin{versiontwo}

DevOps principles, practices, and tools are changing the software industry. However, many industry practitioners, both engineers and managers, are still not aware of how their daily work can be affected by such principles, practices, and tools. By surveying the DevOps literature, we found several implications of DevOps adoption for industry practitioners and for academic researchers. In this section, we lay out such implications as a practical guide to help professionals to adapt to the significant impacts of DevOps in their fields.

Our contribution is to discuss the DevOps implications for each perspective individually: engineers, managers, and researchers. In these implications, readers will find issues they will likely confront, some core concerns they should have, and potential solutions already adopted by the community. We also outline DevOps-related topics that the academic community could exploit in the future. Finally, presenting these implications raises relevant considerations, preparing our discussion of unresolved DevOps challenges in Section~\ref{sec:discussion}.

\end{versiontwo}

\subsection{Implications for engineers}

Based on the reviewed literature, we list DevOps implications affecting how engineers must architect systems, interact with their peers, and even how to adopt processes, such as incident handling.

\myparagraph{Microservices} adopting microservices architecture is recommended in conjunction with adopting continuous delivery~\citesel{shahin2016architecting,balalaie2016microservices}. Loosely coupled and well-encapsulated microservices architecture lead to good system testability and deployability~\citesel{humble2017willwork}. Since microservices do not come without new challenges, we further discuss this topic in Section~\ref{sec:redesign}.

\myparagraph{Cloud services} architecture patterns involving cloud services can assist in the deployment and operation of applications~\citesel{cukier2013patterns}, diminishing the need for dedicated operations teams~\cite{roberts2018serverless}. It is particularly useful in the context of cross-functional teams.  

\myparagraph{Rolling back} several authors state that \devops engineers should be able to roll back applications in case of problems after deployment~\citesel{claps2015journey,callanan2016right}. However, in scenarios with complex integration or database evolution~\citesel{callanan2016right}, rolling back can be a tricky task~\cite{kerzazi2016rollback}. Alternatives such as feature toggle~\cite{hodgson2017toggle} or handling the root causes of problematic releases can avoid rolling back~\cite{microsoft2018devops,kerzazi2016rollback}. 

\myparagraph{Embedded systems and IoT devices} deployment on embedded systems can be hard, especially when only the customer owns and controls the hardware platform~\citesel{lwakatare2016towards}. Therefore, engineers must customize the updating mechanisms for embedded systems and IoT devices.  

\myparagraph{Inhibitors for high-frequency delivery} engineers must be aware of scenarios that could impose barriers to high-frequency delivery.  Updating the software of automation control systems can require a factory to stop its production~\citesel{leppanen2015highways},  which can be costly. Traditional techniques to avoid downtime on deployment, such as Blue Green Deployment~\cite{fowler2010bluegreen} and Canary Release~\cite{sato2014canary}, may not be suitable for critical applications, such as for factories, and medical equipment. Another less critical scenario, such as mobile apps, can wait for a release cycle to publish a new update~\citesel{leppanen2015highways}.

\myparagraph{Testing} although companies recognize the importance of automated testing, they still struggle to implement it~\citesel{olsson2012stairway} fully. It is especially true for user interface tests automation~\citesel{leppanen2015highways}. Other factors that make automated testing complex are hardware availability for load testing and user experiment assessment~\citesel{leppanen2015highways}. Parallelizing test suites can be necessary to reduce testing time~\citesel{neely2013easy}. Tests that seem to fail at random called ``flaky tests,'' are not acceptable~\citesel{neely2013easy}.

\myparagraph{Quality assurance team} quality assurance skills are necessary for locating specific error scenarios and corner cases. However, preserving the quality assurance team separate from the development team is questionable. DevOps and agile practices require a change in the role of quality
assurance teams, or even its elimination~\citesel{bass2018architect}.

\myparagraph{Legacy systems} although it takes a great deal of work to achieve continuous delivery on mainframe platforms, there are successful reports of it~\citesel{humble2017willwork}. Some legacy architectures might not be designed to run automated tests~\citesel{leppanen2015highways}. Nonetheless, teams must be aware that cultural factors, such as managers who say ``This is the way we have always done it''~\citesel{humble2017willwork}, can limit the adoption of continuous delivery more than technical factors.  

\myparagraph{Communication} improving communication among organizational silos is key to \devops adoption~\citesel{humble2011devops}. However, practitioners must be aware that effective interdepartmental communication is still a challenge, especially within remote teams~\citesel{diel2016distributed}. 

\myparagraph{Learning} software professionals must be prepared to learn new tools, focusing on automation~\cite{xebia2016table}. They must also cope with the ever-changing nature of these tools, which implies maintaining heterogeneous environments and migrating technologies~\citesel{shahin2016architecting}. 

\myparagraph{Building the deployment pipeline} the benefits delivered by a deployment pipeline are many. However, engineers must be aware that setting up the infrastructure for continuous deployment can demand a considerable effort~\citesel{leppanen2015highways, neely2013easy}. Breaking down the system into microservices also requires building multiple pipelines. Engineers should not try to build all the continuous delivery ecosystem in a single step: an approach based on continuous improvement is preferable~\citesel{neely2013easy}.

\myparagraph{Pipeline maintenance} pipeline execution generates a lot of artifacts, such as build and logs. Artifacts such as production logs, bug tracks, parameters configuration, and temporary files, must be appropriately archived and removed at some point~\citesel{pang2016maintenance}. Despite its importance to pipeline sustainability, organizations often overlook this maintenance process~\citesel{pang2016maintenance}.

\myparagraph{On-boarding} the automation built for carrying continuous delivery also promotes faster onboarding of new members to the team~\citesel{neely2013easy}.

\myparagraph{Incident handling} software developers must be educated in software security and must cope with incident handling~\citesel{jaatun2018incident}, bug tracking, systems failure, or take over primary responsibility for this activity.

\myparagraph{Coding for stability and security} although software stability and security are traditional operational concerns, in a \devops context such non-functional requirements must be leveraged by software implementation, including the code for deployment automation and a consideration of the possibility of failures and delays in the underlying infrastructure~\citesel{rahman2018defective, li2018api, zhu2015frequency}. 

\subsection{Implications for managers}

Based on the reviewed literature, we list implications related to how managers must face the DevOps phenomenon: required management and cultural paradigms, training people, structuring and assessing the DevOps-adoption process, as well the expected outcomes from this process.

\myparagraph{Adoption of lean principles} since \devops is based on lean principles~\citesel{claps2015journey}, organizations eager for \devops adoption should take a step back and learn them~\cite{poppendieck2006lean}. In particular, Kim recommends~\cite{kim2012ways}: 1) mapping the value stream for optimizing global system performance, rather than for local optimization; 2) amplifying continuous feedback loops to support necessary corrections; and 3) improving daily work through a culture promoting frequent experimentation, risk-taking, learning from mistakes, and knowing that practice and repetition are prerequisites to mastery.

\myparagraph{\devops adoption} how to deploy \devops in an organization is a critical question for managers. However, the lack of a consensual definition of DevOps is a reason for making it a hard decision. Many studies explore this subject, organizations still struggle with it, and we discuss it in more depth in Section~\ref{sec:devosp-adoption-approaches}. 

\myparagraph{Assessment} an organization should be able to measure the success of a \devops adoption process. Nonetheless, any top-down imposition of a metric-based evaluation must be used with care. If personal evaluation depends on such metrics, engineers can focus on producing good numbers rather than improving the software process~\cite{microsoft2018devops}. We further discuss how to assess the quality of DevOps practices in Section~\ref{sec:assessing-devops}.

\myparagraph{Training}  \devops demands additional technical skills from software professionals. 
Developers must acquire skills from operators and vice versa. It is necessary to conduct training in using not only tools but also in using \devops concepts to keep pace with tooling's rapid pace evolution~\citesel{shahin2016architecting}. Top management sponsorship for training is often necessary for many contexts. 

\begin{versiontwo}
\myparagraph{Job titles} Defining job titles impacts hiring and training. Although the DevOps movement emerged to approximate developers and operators, nowadays, the industry adopts the role of DevOps engineer, which executes tasks mostly linked to scripting automation and CI/CD practices~\citesel{hussain2017zealand}. However, this role blurs with other ones, such as release engineer and build engineer. There is no consensus in industry and academia when defining these roles. By analyzing hiring ads, Hussain \emph{et al.} found that DevOps positions usually do not impose build and release management as attributions~\citesel{hussain2017zealand}, whereas Kerzazi and Adams found that these three roles share common activities~\cite{kerzazi2016whoneeds}. The challenge for managers to define job titles is also related to how to structure development and operations teams, which we discuss in Section~\ref{sec:devosp-adoption-approaches}.
\end{versiontwo}

\myparagraph{Culture} Humble stated that, typically, the obstacle to continuous delivery adoption is not the skill level of individual employees, but failures at the management and leadership level~\citesel{humble2017willwork}. High-performing organizations strive, at all levels, towards continuous improvement, rather than treating workers as fungible ``resources'' that should merely execute tasks as efficiently as possible~\citesel{humble2017willwork}. Top and low-level management are responsible for creating an environment in which failure is allowed, and people seek continuous improvement.

\myparagraph{Increase in delivery throughput} Neely and Stolt reported that the delivery throughput per developer in their company increased with the adoption of continuous delivery and defect rate became more predictable~\citesel{neely2013easy}. Siqueira~\emph{et al.} also reported how, after investments in \devops and continuous delivery,  the number of releases per semester remained the same even after a considerable decrease in the number of team members~\citesel{siqueira2018gov}.

\myparagraph{Building trust} when an organization builds software for other large organization, especially the government, it is common for the relationship between the contractor and contracted to be dominated by mistrust, which leads to cumbersome development processes. However, Siqueira~\emph{et al.} advocate, based on their experience, that continuous delivery leverages a trust relationship among the contractor and contracted in software development, even in governmental projects~\citesel{siqueira2018gov}.

\myparagraph{Building for the government} although building software for government involves more bureaucratic processes, requirements and prioritization can often change due to political reasons~\citesel{siqueira2018gov}, which leads to the need to shorten the release cycle by adopting agile and DevOps practices in the governmental scenario.

\subsection{Implications for researchers}

Based on the reviewed literature, we list DevOps-related open topics that the academic community could exploit in future research.

\myparagraph{Software Architecture} some authors explore software design in the context of \devops, continuous delivery, and continuous deployment~\citesel{shahin2016architecting,woods2016view,brown2016microservices}. However, engineers may still struggle with this in practice, since achieving the desired architecture can be infeasible in a single first \devops project. Researchers should investigate transitioning strategies to adopt practices and architectural changes. 

\myparagraph{Education} \devops adoption requires more skilled software engineers. A broader investigation is needed to explore how to teach operations skills to developers and vice-versa, as well as to introduce software operation topics in software engineering courses~\citesel{christensen2016teaching,chung2016ksa}. We discuss this in more detail in Section~\ref{sec:qualify}. 

\myparagraph{Embedded systems and IoT} continuously delivering new software versions in embedded systems is still a hard question~\citesel{lwakatare2016towards,olsson2012stairway}. How to effectively adopt \devops in an IoT context is an open research question to be investigated. 

\myparagraph{Compliance} \devops provides both benefits and obstacles to regulatory compliance~\citesel{laukkarinen2017medical,humble2011devops,humble2017willwork}. Research should be done to detect scenarios in which a high-frequency delivery is indeed not welcome or how one can demonstrate that automated deployment can help organizations comply with regulations, as stated by Humble~\citesel{humble2017willwork}. Moreover, some practitioners advocate that engineers must have unrestricted access to production data~\cite{chef2015kungfu,netflix2018devops}, which can be controversial in specific environments, such as financial systems. 

\myparagraph{Security} high-frequency of continuous delivery poses questions about the security in a \devops context~\citesel{yasar2016security,jaatun2017security}. However, \devops can also bring benefits for security. This dual relationship generates a dichotomy that should be further investigated. 

\myparagraph{Testing large-scale distributed systems} some errors in large-scale distributed systems can be hard to reproduce in testing environments~\citesel{gray2006conversation}. The chaos engineering approach advocates running experiments in production~\citesel{basiri2016chaos}. This is an incipient topic in the research literature. 

\myparagraph{Quantitative assessment metrics} few works use adequate quantitative metrics for assessing \devops as is the case of the Puppet State of \devops Reports~\cite{puppet2014devops}. The investigation and use of such quantitative metrics should be intensified in related research.  

\myparagraph{Deployment approach} two automated deployment approaches are the container-based one~\citesel{kang2016docker} and the deployment scripts written in domain-specific languages such as Chef. Investigating how complementary or exclusive these approaches are can help engineers choose the best toolset. 

\myparagraph{Improving interdepartmental communication} how to effectively enable interdepartmental communication is a challenge, especially for distributed organizations~\citesel{diel2016distributed}. Better communication does not merely mean more communication since too much communication, and an excess of meetings can negatively impact productivity. 
 
\myparagraph{Adoption strategies} there are still many open questions about how organizations should adopt \devops. It is stated that \devops adoption requires top-management support~\citesel{claps2015journey,olsson2012stairway}. Sometimes it does not happen in the first moment, and a ``guerrilla'' strategy can take place (i.e., acting outside the company standard procedures). Moreover, arguments to encourage \devops adoption can differ from engineers to managers~\citesel{neely2013easy}. In such context, researchers can support \devops adoption by providing a further investigation into balancing top-management support and guerrilla strategies; studying arguments to convince managers; providing guidelines about which \devops-adoption strategy to choose based on organizational characteristics.

\myparagraph{Investigation involving multiple organizations} since most empirical research is conducted with data from a single organization~\citesel{nybom2016mixing,nybom2016mixing,claps2015journey,diel2016distributed}, any conclusion from these works deserves further investigation. Surveying more companies, as done by Leppanen \emph{et al.}~\citesel{leppanen2015highways}, can strengthen or weaken previous findings. 

\myparagraph{Other research topics} Claps \emph{et al.} list other technical and social challenges for continuous deployment adoption, such as product marketing, team coordination, customer adoption, feature discovery, plugin management, cross-product dependencies, and scaling CI tools~\citesel{claps2015journey}. Olsson \emph{et al.} list among the challenges in continuous delivery adoption: coordination of supplier integration, business models, and difficulty in overviewing projects status~\citesel{olsson2012stairway}.

\section{Unresolved Challenges}
\label{sec:discussion}

Throughout this survey, we have shown that overviewing \devops concepts and tools raises some issues to debate. The previous section also listed some not-so-straightforward implications for practitioners and researchers. Therefore, based on the surveyed literature and additional sources of knowledge, we further discuss in this section some of the \devops challenges faced by managers, engineers, and researchers that are not thoroughly handled by the current state-of-the-art.  

\subsection{How to re-design systems toward continuous delivery}
\label{sec:redesign}

Highly coupled monolithic architectures are obstacles to effective continuous delivery. Complex dependency management of software components and teams is imposed to the deployment pipeline~\citesel{shahin2016architecting}.
An essential principle for successfully adopting and implementing continuous delivery is an architecture composed of small and independently deployable units, also called \emph{microservices} ~\citesel{shahin2016architecting}. In such architectural style, services interact through the network and are built around business capabilities~\cite{fowler2014microservices}.

Conceiving of an application architecture with loosely coupled and well-encapsulated microservices guarantees two architectural attributes required by continuous delivery: \emph{testability} and \emph{deployability}~\citesel{humble2017willwork}. If each microservice has its test suite, this reduces the blocking of deployment due to long-running tests~\citesel{leppanen2015highways}.

Whereas some authors say microservices facilitate effective implementation of \devops~\citesel{balalaie2016microservices}, others say microservices require \devops~\citesel{ebert2016devops}, since deployment automation minimizes the overhead to manage a significant number of microservices.
However, adopting microservices comes with several challenges. First,  there is heterogeneity in non-functional patterns such as ``startup scripts, configuration files, administration endpoints, and logging locations''~\citesel{callanan2016right}. Technological heterogeneity can be a productivity barrier for newcomers in the team. Second, microservices must be deployed to production with the same set of versions used for integration tests~\citesel{brown2016microservices}. These challenges can be overcome by adopting a minimum set of microservices standards across the organization~\citesel{callanan2016right}.

Common microservice management patterns associated with \devops are: one deployment pipeline per microservice; log aggregator; service registry; correlation ID's~\citesel{brown2016microservices}; and segregation of source code, configuration, and environment specification~\citesel{balalaie2016microservices}. 
Other complimentary patterns are strangler application~\cite{fowler2004strangler}\citesel{humble2017willwork,brown2016microservices}, load balancer, consumer-driven contracts~\citesel{balalaie2016microservices}, circuit breaker~\cite{nygard2009release}, backwards compatibility, and versioned APIs~\citesel{humble2017willwork}. However, the last one is controversial, and some communities usually do not recommend microservice versioning~\citesel{balalaie2016microservices}. 

In highly regulated environments, the proper use of microservices can constrain unfriendly regulations to specific system modules and people~\citesel{humble2017willwork}. Additionally, the usage of a platform-as-a-service (PaaS) can automate much of the compliance checking, a model used, for example, by the U.S. federal government~\citesel{humble2017willwork}. 
PaaS is also recommended for its operational simplicity, like other cloud services, such as storage services, asynchronous processing with queues, email delivery, and real-time user monitoring~\citesel{cukier2013patterns}.

Finally, although reusability is a historical goal of software engineering~\cite{mcilroy1968massproduced}, Shahin \emph{et al.} recommend engineers not to focus too much on it~\citesel{shahin2016architecting}. It brings coupling, which is a huge bottleneck to continuous delivery~\citesel{shahin2016architecting}.  Therefore, each team should discuss the trade-off between reusability and independence from other components, services, and teams.

\subsection{How to deploy \devops in an organization}
\label{sec:devosp-adoption-approaches}

A hard question not yet fully answered by the literature is how -- perhaps whether -- an organization should be restructured to adopt \devops. We believe the literature is incomplete and even contradictory regarding this subject. We discuss a few studies that have evaluated this matter.

The seminal paper of Humble and Molesky about \devops presents three scenarios. First, preserving the structures of development and operations departments,  \devops is led by human collaboration among developers and operators, with operators attending agile ceremonies and developers contributing to incident solving~\citesel{humble2011devops,jaatun2018incident}. 
Alternatively, product teams, also called cross-functional teams, effectively incorporate operators. Finally, in a third scenario, one product team is composed of only operators offering support services (continuous integration, monitoring services) to the entire organization.

Similarly, Nybom \emph{et al.} also present three distinct scenarios to \devops adoption~\citesel{nybom2016mixing}: \emph{i)} collaboration among development and operations departments; \emph{ii)} cross-functional teams; and \emph{iii)} creation of ``\devops teams.'' These approaches are summarized in the Table~\ref{tab:adoption-approaches} and discussed in detail in the following paragraphs.

{
\small

\begin{table}[ht]
\centering
\caption{\devops adoption approaches}
\label{tab:adoption-approaches}
\begin{tabular}{l} \hline

\textbf{Collaborating departments} \\
Development and operations departments collaborate closely.\\
It implies overlapping of developers and operators responsibilities.\\
\emph{Downside:} new responsibilities can be unclear for employees.\\
\hline

\textbf{Cross-functional team} \\
The product team is responsible for deploying and operating (\emph{You built it, you run it}).\\
Recommended by Amazon and Facebook.\\
\emph{Downside:} requires more skilled engineers.\\
\hline 

\textbf{\devops team} \\
Acts as a bridge between developers and operators. \\
It is better accepted when it is a temporary strategy for cultural transformation.\\
\emph{Downside:} risk of creating a third silo.\\
\hline

\end{tabular}
\end{table}

% end small
}

In the first approach, the responsibilities of developers overlap with operators~\citesel{nybom2016mixing,claps2015journey,shahin2016architecting,diel2016distributed}.  High risk for adopting this strategy occurs when new responsibilities do not become evident in the organization~\citesel{nybom2016mixing,claps2015journey}, which could lead to frictions among developers and operators~\citesel{chen2015huge}. This is especially true when delivery automation is not prioritized~\citesel{nybom2016mixing}. 

Cross-functional teams resemble the ``whole team'' practice from XP~\cite{beck2004xp}, and seems to prevail in literature~\citesel{feitelson2013facebook,gray2006conversation,cukier2013patterns,balalaie2016microservices}.  This structure appeals to ``T-shaped'' professionals, who have expertise in few fields and basic skills in multiple correlated areas~\citesel{debois2011revolution}. In this model, at least one team member must master operations skills. From this perspective, the so-called ``\devops engineer''~\citesel{hussain2017zealand}, also called the ``full stack engineer''~\citesel{hussain2017zealand},  is known as a developer with operations skills.

One could wonder whether it makes sense to talk about developers and operators collaboration in cross-functional teams context. Adopting cross-functional teams may just be a matter of moving operators to product teams. However, this is not trivial, considering that large organizations usually have only a small pool of operators for several development teams~\citesel{nybom2016mixing}. Therefore, companies must hire people with both development and operations skills or train some of their developers in operations skills. 
Nevertheless, pressuring developers to learn operations can be ineffective, as they are already overwhelmed with other skills they need to learn. Transforming development teams into product teams has still another major cultural shift, as developers become responsible for 24/7 service support~\citesel{shahin2016architecting, debois2011revolution}. Even when services should be designed to remain available without human interaction~\cite{hamilton2007designing}. 
An extra challenge for cross-functional teams is guaranteeing that specialists interact with their peer groups to share novelties in the field~\citesel{debois2011revolution}. Spotify, for example, achieves this through inter-team ``guilds,'' which are communities of common interest within the organization~\cite{spotify2014culture}.

Assigning a ``\devops team'' as a bridge between developers and operators~\citesel{nybom2016mixing} has become a trend~\cite{puppet2014devops}. However, it is not clear what precisely ``\devops teams'' are, and how they differ from regular operations teams~\citesel{nybom2016mixing}\cite{puppet2014devops}.  This approach is criticized by Humble, who considers that potentially creating a new silo is not the best policy to handle the \devops principle of breaking down walls between silos~\cite{humble2012team}. 
Skelton and Pais present some variations, called ``\devops topologies,'' and also \devops anti-patterns~\cite{skelton2013topologies}. They argue that, although the \devops team silo is an anti-pattern, it is acceptable when it is temporary and focused on sharing development practices to operations staff and operations concerns to developers. Siqueira~\emph{et al.} report how a \devops team with senior developers and rotating roles among members was a key decision to construct and maintain a continuous delivery process~\citesel{siqueira2018gov}.

There is yet the strategy adopted by Google, called Site Reliability Engineering (SRE)~\cite{beyer2016site}, which involves the evolution of the operations engineer role. In addition to product teams, Google has SRE teams responsible for product reliability engineering, what includes performance, availability, monitoring, and emergency response. SRE teams have an upper limit of using 50 percent of their time on operational tasks because they must allocate at least 50 percent of their time increasing product reliability through engineering and development activities.

One should also question the impact of deployment automation on operators' roles. Chen reported that, after continuous delivery adoption, operations engineers only needed to click a button to release new versions~\citesel{chen2015huge}. One should question the role of an engineer in such a context.
Moreover, if software updates are adequately broken down into small increments, and delivery automation turns deployment in a ``non-event''~\citesel{humble2011devops}, practices such as 
``development and operations teams should celebrate successful releases together''~\citesel{humble2011devops}
can be perceived as contradictory.

Therefore, if operations provide support services (continuous integration, monitoring)~\citesel{humble2011devops} and product reliability is appropriately designed, product teams can easily deploy and operate their services. Thus, one can even replace operations with third-party cloud services, as done by Cukier~\citesel{cukier2013patterns}, and achieve an organizational structure known as ``NoOps''. 
Some blog posts promote  NoOps concepts, but the literature has not yet covered this subject. One possible definition is: ``NoOps (no operations) is the concept that an IT environment can become so automated and abstracted from the underlying infrastructure that there is no need for a dedicated team to manage software in-house''~\cite{rouse2015noops}. Cross-functional teams can embrace NoOps, with a particular focus on employing cloud technology that automates much of the operational work, rather than reallocating operators or teaching developers all the operational work.

\subsection{How to assess the quality of \devops practices in organizations}
\label{sec:assessing-devops}

Some researchers have proposed maturity models for \devops adoption and continuous delivery adoption~\cite{atlassian2017maturity,beal2015maturity,leppanen2015highways}. Feijter~\emph{et al.}, for example, provide maturity \devops models to help organizations measure their current maturity level and to determine areas that require additional investment~\citesel{feijter2018maturity}. However, none of these maturity models are widely adopted by industry. Indeed, most organizations have almost no visibility or reliable measurement of their software-delivery practices~\citesel{forsgren2018metrics}.

Forsgren and Kersten claim that both survey data and system data~\citesel{forsgren2018metrics} should measure \devops transformations. System data can provide a continuous flow of information, although setting up the aggregation of metrics from different sources can be challenging. Survey data provides a holistic view of out-of-the system issues, such as culture, job satisfaction, and burnout, and can even point to problems on system-data collection. 

Whether survey or system data, Forsgren and Kersten still make clear that if results are used to punish teams, data collection will be unreliable~\citesel{forsgren2018metrics}, as also reinforced by Brown from Microsoft~\cite{microsoft2018devops}. Kua cautions that inappropriate use of metrics can lead to undesired behavior and even divert the organization from its goals~\cite{kua2013metrics}. Kua also compiles guidelines for better usage of metrics by: explicitly linking metrics to goals, favoring trends over absolute numbers, tracking shorter periods, and changing metrics when they do not drive change anymore.

Some examples of metrics based on survey data used on the State of DevOps Reports~\cite{puppet2014devops, puppet2016devops} are: IT performance as a function of time from commit to deployment, frequency of deployment, and recovery time. Time spent on unplanned work and rework. Typology of organizational culture (pathological, bureaucratic or generative) based on climate for learning, job satisfaction, developers and operators having win-win relationships, usage of version control, automated testing.
Employee engagement, based on the Net Promoter Score, indicating how likely employees are to
recommend the company products and services.

Snyder and Curtis use metrics to evaluate \devops in a specific company~\citesel{snyder2018analytics}. Metrics such as productivity, defect ratio, dollars spent, cycle time release, and build count were aggregated in a ``total quality index''. More importantly, the aggregation of data from different silos was used to align organizational efforts. The authors observed a higher increase in the total quality index of agile and \devops teams, which helped executives justify investments in the agile-\devops transformation.

There is a severe lack of industry consensus on how to measure software delivery~\citesel{forsgren2018metrics}, and designing a good survey requires some expertise. Thus, researchers should provide standards for collecting and analyzing evaluation metrics for the software delivery process, as done by Feijter~\emph{et al.}~\citesel{feijter2018maturity}. In the health sector, this model is mature, wherein researchers instrument therapists with standard protocols for diagnoses purposes. Nonetheless, Forsgren and Kersten provide some general guidelines for survey application: it should be limited to 20 to 25 minutes and applied every four to six months~\citesel{forsgren2018metrics}. Forsgren and Kersten also advise avoiding surveying management and executives, since they tend to overestimate the maturity of their organizations.

Researchers can also run the surveys across multiple organizations as done in the State of DevOps Reports. The results of these questionnaires can help practitioners in orienting their careers and help organizations to self-evaluate their performance by comparing themselves to their peers.

\subsection{How to qualify engineers for \devops practice}
\label{sec:qualify}

More than one-hundred DevOps tools are available~\cite{xebia2016table}, and they are continuously evolving~\citesel{shahin2016architecting}. Professionals must be able to choose which tools to study in depth, which tools to have some familiarity with,  which tools to test, and which ones to ignore.  Such choices must be based on a conceptual analysis of personal and organizational demands and perspectives. Our conceptual maps (Section~\ref{sec:concepts}) and our table of major \devops tools (Table~\ref{tab:tools}) may guide engineers to prioritize the improvement of their professional skills.

Care is needed when managers assume that employees lack competence. When questioned ``Where do you get Netflix's amazing employees from?'', a Netflix architect replied: ``I get them from you!''~\citesel{humble2017willwork}. The lesson here is that a culture of autonomy and responsibility~\citesel{gray2006conversation,feitelson2013facebook} is an important motivating factor for employees continuous learning.

In software engineering courses, operations skills, often neglected in college education,  become mandatory.  However, elaborate effective \devops programs are challenging. Exercises that demand students build distributed systems with adequate availability, scalability, and performance require not only some infrastructure, but also efficient correction of such exercises with automated checks in submissions~\citesel{christensen2016teaching}. 
Although there are stable platforms for auto-grading coding assignments (\url{autolab.github.io})~\cite{dejan2011autograding}, evaluating infrastructure code and non-functional concerns are harder to automate.
Christensen used Docker in the classroom to develop a small set of scripts to automate some tasks in executing and assessing submissions~\citesel{christensen2016teaching}. However, he did not achieve automation of exercise correction due to the nontrivial task of setting up a multi-server environment to detect errors in students' submissions. Bai~\emph{et al.} automate the generation of assessment reports based on patterns of tasks, commits, branches, tests, and the source code itself~\citesel{bai2018feedback}.

Knowledge in software engineering is always evolving, and any software engineer must also have self-learning skills to overcome daily basis challenges. Social interactions on Internet forums are a vital source of information in the daily activities of engineers. Believing that it can be enhanced, Magoutis \emph{et al.} built a social network that helps \devops engineers analyze results of past deployment executions as well as communicate and exchange ideas to better understand trade-offs in the space of deployment solutions~\citesel{magoutis2015social}. The work of Magoutis \emph{et al.} show evidence that more can be done to support engineers' self-learning in their daily activities.

\section{Limitations of this study}
\label{sec:limitations}

Given the vast amount of work on \devops, it was not feasible to conduct an exhaustive search in all available sources. We have not considered academic workshops, and we mentioned only a few books in the field. It would not be practical to try to cover every existing work on \devops and to condense it all in this single survey. However, by focusing on journal and conference papers, and by applying the snowballing process, we aimed to cover the primary literature in the field. We decided not to expand the query string to avoid artificially favoring some of the \devops concepts over others.  However, the appearance of the keyword ``\devops'' was not mandatory in the snowballing process so that we could select vital work tackling highly related subjects such as continuous delivery. 

The choice of core and relevant papers may suffer from subjective bias. To reduce this bias,
at least two authors conducted the selection process and the classification of papers. 
Senior researchers also oversaw the whole process.
Our study also suffers the publication bias, which is the propensity of researchers and practitioners to report (and for referees to accept) more about positive results than negative results. We did not find, for example, a paper reporting a case of failed DevOps adoption. We did not address this issue in this study.

\section{Conclusions}
\label{sec:conclusions}

In this survey, we have discussed \devops concepts and challenges presented in the literature. By associating these concepts with tools, we contributed to supporting practitioners in choosing a proper toolset. This paper also aides IT professionals by presenting in a systematic way the most relevant concepts, tools, and implications, associated with the professional perspectives of researchers, managers, and engineers -- including developers and operators. We hope our reader can now better understand the impact of \devops on daily activities based on each profile.

Two pillars of \devops are 1) human collaboration across departments and 2) automation. We found that technical issues regarding delivery automation are more consolidated, with only a few minor controversies in the literature. On the other hand, there is no consensus on how to effectively empower collaboration among departments, and there are only a few tools for tackling this issue.

A technical topic highly related to \devops revealed by our sources was microservice architecture. Some technical conflicting points are: there is no consensus on whether microservices should be versioned; some practitioners see configuration management tools and containerization as complementary approaches, but others understand them as competitors; maximizing reusability of software components and services may not be the better strategy; rolling back is usually seen as a \devops practice, but some practitioners tend to favor feature toggles to avoid these rollbacks.

Structuring an organization to introduce \devops is now a major concern. There are three alternative structures: 1) preserving collaborating departments, 2) building cross-functional teams, and 3) having \devops teams. Each one of these strategies has its challenges and, by following the current literature, it is not possible to define the best strategy for a given scenario. \vtwo{Therefore, exploring these and other structures in real organizations and how these organizations handle the ``DevOps role'' are promising topics for future research.  However, whatever the structure an organization adopts}, it is clear that the \devops movement has irreversibly blurred the frontier between developers and operators, even for organizations that have not yet fully embraced \devops.

\section*{Acknowledgements}

{\small
We thank the Brazilian Ministry of Citizenship, via the TAIS project, and the Brazilian Federal Data Processing Service (Serpro) for the support.
This research is also part of the INCT of the Future Internet funded by CNPq proc. 465446/2014-0, Coordena\c{c}\~ao de Aperfei\c{c}oamento de Pessoal de N\'{\i}vel Superior - Brasil (CAPES) - Finance Code 001, FAPESP proc. 14/50937-1, and FAPESP proc. 15/24485-9.
}
% Bibliography
\bibliographystylesel{ACM-Reference-Format}
\bibliographysel{selected-studies}
\bibliographystyle{ACM-Reference-Format}
\bibliography{references}

\end{document}